\documentclass[preprintnumbers,letterpaper,aps,prd,nofootinbib,
twocolumn
]{revtex4-1}
\usepackage{bm,graphicx,dcolumn,epstopdf,epsf, latexsym,mathbbol, amssymb,amsmath,color,slashed, mathrsfs,mathcomp,simplewick,float}
\usepackage{palatino} \usepackage{mathpazo} 
\usepackage{amsfonts}
\usepackage[center]{subfigure}
\usepackage{multirow}
\usepackage{makecell}
\usepackage[colorlinks,linkcolor=red,citecolor=blue,urlcolor=blue]{hyperref}

\usepackage{anyfontsize} 

\usepackage{silence}
\newcommand{\Eqref}[1]{(\ref{#1})}
\newcommand{\brac}[1]{\left(#1 \right)}

\allowdisplaybreaks
\newcommand{\bq}{\begin{equation}}
\newcommand{\eq}{\end{equation}}
\newcommand{\bqn}{\begin{eqnarray}}
\newcommand{\eqn}{\end{eqnarray}}
\newcommand{\nb}{\nonumber}
\newcommand{\lb}{\label}

\begin{document}

	\title{Orbital mechanics and quasiperiodic oscillation resonances of black holes in Einstein-{\AE}ther theory}
	
	\author{Mustapha Azreg-A\"{i}nou${}^{a}$}
	\email{azreg@baskent.edu.tr}
	
	\author{Zihang Chen${}^{b}$}
	\email{MAT1809527@xmu.edu.my}
	
	\author{Bojun Deng${}^{b}$}
	\email{MAT1809529@xmu.edu.my}
	
	\author{Mubasher Jamil${}^{c,d}$ }
	\email{mjamil@zjut.edu.cn; Corresponding author}
	
	\author{Tao Zhu${}^{c, e}$}
	\email{zhut05@zjut.edu.cn}
	
	\author{Qiang Wu${}^{c}$}
	\email{wuq@zjut.edu.cn}
	
	%
	
	\author{Yen-Kheng Lim${}^{b}$}
	\email{yenkheng.lim@gmail.com}
	
	\affiliation{${}^{a}$ Ba\c{s}kent University, Engineering Faculty, Ba\u{g}l{\i}ca Campus, 06790-Ankara, Turkey}
	
	\affiliation{${}^{b}$ Department of Mathematics, Xiamen University Malaysia, 43900 Sepang, Malaysia}
	
	\affiliation{${}^{c}$ Institute for Theoretical Physics and Cosmology, Zhejiang University of Technology, Hangzhou, 310023, China}
	
	\affiliation{${}^{d}$ School of Natural Sciences, National University of Sciences and Technology, Islamabad, 44000, Pakistan}
	
	\affiliation{${}^{e}$ United Center for Gravitational Wave Physics (UCGWP), Zhejiang University of Technology, Hangzhou, 310023, China}

	\date{\today}

	\begin{abstract}
		
		In this paper, we study the motion of test particles around two exact charged black-hole solutions in Einstein-{\AE}ther theory. Specifically, we first consider the quasi-periodic oscillations (QPOs) and their resonances generated by the particle moving in the Einstein-{\AE}ther  black hole and then turn to study the periodic orbits of the massive particles. For QPOs, we drop the usually adopted assumptions $\nu_U=\nu_\theta$, $\nu_L=\nu_r$, and $\nu_U/\nu_L=3/2$ with $\nu_U$ ($\nu_L$) and $\nu_r$ ($\nu_\theta$) being the upper (lower) frequency of QPOs and radial (vertical) epicyclic frequency of the orbiting particles, respectively. Instead, we put-forward a new working ansatz for which the Keplerian radius is much closer to that of the innermost stable circular orbit and  explore in detail the effects of the {\ae}ther field on the frequencies of QPOs. We then realize good curves for the frequencies of QPOs, which fit to data of three microquasars very well by ignoring any effects of rotation and magnetic fields.  The innermost stable circular orbits (isco) of timelike particles are also analyzed and we find the isco radius increases with increasing $c_{13}$ for the first type black hole while decreases with increasing $c_{14}$ for the second one. We also obtain several periodic orbits and find that they share similar taxonomy schemes as the periodic equatorial orbits in the Schwarzschild/Kerr metrics, in addition to exact solutions for certain choices of the Einstein-\AE ther parameters. The equations for null geodesics are also briefly considered, where we study circular photon orbits and bending angles for gravitational lensing.
		
		\end{abstract}
	
	
	\maketitle

	\section{Introduction\label{secint}}

	Numerous observable candidates of astrophysical black holes are never isolated systems. In fact, they are the center stage of immense astrophysical activities such as swirling high speed dust and gases in the accretion disks, astrophysical X and $\gamma-$rays emitting  jets,  and stellar disruptions and collisions \cite{fabian}. By analyzing the dynamics of the particles, which follows geodesics around the black holes, one is able to infer the geometric structure of the spacetime with extreme curvature. In such investigations, the motion of particles in the circular orbits around black holes have foremost relevance. In the particular case of a Schwarzschild black hole, the light ring which forms by the photon's unstable circular orbits and the innermost stable circular orbit (isco) for massive particles have the radii, $3M$ and $6M$, respectively. For the spinning black hole, the size of these orbits become larger or smaller depending on the value of the spin and the angular momentum of the moving particles \cite{ab}. The light ring surrounding the black hole forms a shadow which can become more disoriented if the black hole's spin is increased or if the values of the free parameters appearing in the spacetime metric are modified. Thus the nearest orbits of photons or massive particles provide deeper information about the geometrical structure and the physical processes happening in the strong gravity regions of the black holes. In this context, the recent discovery of the shadow of the central supermassive black hole at the center of M87 galaxy by EHT Collaboration provided necessary data to test other modified gravity black holes and constrain their parameters \cite{EHT}. In the literature, the circular orbits around various kinds of black holes in various gravitational setups have been studied extensively \cite{lit}.
	
	Among numerous astrophysical events, the QPOs are a very common phenomena in the X-ray power density spectra of stellar-mass black holes. The frequency of QPOs can be related to the matter orbiting near the isco of the black hole, thus is very sensitive to geometrical structure of the black hole spacetime. In particular,  the appearance of two peaks at 300 Hz and 450 Hz in the X-ray power density spectra of Galactic microquasars, representing possible occurrence of a lower QPO and of an upper QPO in a ratio of 3 to 2, has stimulated a lot of theoretical works to explain the value of the 3/2-ratio. Some theoretical models, including parametric resonance, forced resonance and Keplerian resonance have been proposed. Therefore, the study of QPOs not only help us understand the physical processes in black hole mechanics, but also importantly, provides a powerful approach to explore the nature of the black hole spacetime in the regime of strong gravity.
	
	Another step towards probing the geometry around a black hole is via particle orbits, the characteristics of periodic orbits may serve to illuminate the underlying structure of all orbits, particularly bound ones \cite{Levin:2008mq}. In Refs.~\cite{Levin:2008mq,Levin:2009sk}, Levin et al. demonstrated how orbits around Schwarzschild and Kerr black holes can be classified under a taxonomy of its periodic orbits, and that any aperiodic orbits may be approximated by a periodic one up to arbitrary high accuracy. The periodic orbits around black hole spacetime can also play an essential role in the study of the gravitational waves, since the periodic orbits can act as successive orbit transition states  of two merging black holes with extreme mass ratio in their inspiral stage \cite{GWs}. In the literature, the study of the periodic orbits have also been carried out in several black holes, see \cite{periodic} and references therein.
	
	
	On the other hand, the phenomenological study of the Einstein-{\AE}ther theory have attracted a lot of attentions. The Einstein-{\AE}ther theory is a generally-covariant theory of gravity which violates the Lorentz symmetry locally. In this theory, the presence of the {\ae}ther field (a timelike vector field) defines a preferred timelike direction that violates the Lorentzian symmetry unlike general relativity (GR) \cite{davi,d1}.  Various implications of this kind of {\ae}ther field have been explored in cosmological contexts such as cosmological perturbations \cite{ li, bat}, the effects on the generation and propagation of gravitational waves \cite{Zhang:2019iim}, and shadow of black holes \cite{Zhu:2019ura}, etc. Moreover the astrophysical constraints on the coupling parameters of the theory are also studied with the data of the gravitational wave events GW170817 and GRB 170817A \cite{aw}.  Recently, two static, charged, and spherical symmetric black hole solutions have been found in the Einstein-{\AE}ther theory with two specific combination of the coupling constants \cite{ding_charged_2015}. Another spherically symmetric black hole solution for a class of coupling constants has also been explored by using numerical calculation \cite{Eling:2006ec} and its analytical representation in the polynomial form has been used in the study of the quasi-normal modes in the Einstein-{\AE}ther theory \cite{Konoplya:2006rv, Konoplya:2006ar}. The laws of black hole thermodynamics and the analysis of cosmic censorship conjecture involving the existence of universal horizons during gravitational collapse have been investigated in \cite{af}.
	
	Therefore, with  the  above  mentioned  motivations, it is of great interest to explore the effects of the {\ae}ther field on properties of the black holes and its astrophysical implications. For this purpose, in this paper, we study the motion of test particles around two exact black holes in the Einstein-{\AE}ther theory. Specifically, we first consider the QPOs generated by the motion of test particles in the Einstein-{\AE}ther black hole and then turn to study the periodic orbits of the massive particles. The null geodesics for the motion of massless particles and gravitational lensing effects by the  Einstein-{\AE}ther black hole has also been considered. As we mentioned, the QPOs and periodic orbits of particles around the black hole can be related to the electromagnetic and gravitational spectra of black holes, therefore, our study here could lead to deeper insights into the observations of X-ray sources and gravitational wave events. Since there is no rotating black hole solution to Einstein-{\AE}ther theory, our investigations and analyses will concern only the two static charged black hole solutions. We shall ignore other important physical aspects such as particle spins, magnetic fields, back-reaction or self-force effects as well as the frame-dragging effects.
		
	The paper is organized as follows. In Sec.~\ref{seceae}, we provide a brief review of Einstein-{\AE}ether theory and the two charged static black hole solutions within this theory. In Sec.~\ref{sececos}, we determine generic expressions for the radial and vertical QPOs then apply them to Einstein-{\AE}ether theory. We will also discuss the energy radiated by a charged particle. In a further application of QPOS, we show how to obtain a good and complete curve fits to the data of three microquasars. In Sec.~\ref{secgeod}, we generally discuss the geodesic equations and their solutions. In particular, we solve the equations of path and the bending angle obeyed by massive particles. Sec.~\ref{secmb} is devoted to the study of marginally bound orbits and the innermost stable circular orbits. We also calculate the energy and angular momentum carried by the particles in these orbits. In Sec.~\ref{secpo}, we discuss the periodic orbits and Sec. ~\ref{nullgeo} deals with null geodesics while in Sec.~\ref{seccon} we explore the gravitational lensing phenomenon for Einstein-\AE ther black holes. Finally we present a conclusion in the terminal section.

	\section{Black hole solutions in Einstein-\AE ther theory\label{seceae}}

	In this section, we present a brief review of the black hole solutions in the Einstein-{\AE}ther theory.
	
	\subsection{Field Equations }
	
	The Einstein-{\AE}ther theory involves, in addition to the spacetime metric tensor field $g_{\mu\nu}$, a dynamical, unit timelike {\ae}ther field $w^{\alpha}$ (it is also called {\ae}ther four velocity) \cite{jacobson_gravity_2001, foster_radiation_2006, garfinkle_numerical_2007, jacobson_einstein-aether_2008}. Like the metric, and unlike other classical fields, the {\ae}ther field $w^{\alpha}$ cannot vanish anywhere, so it breaks local Lorentz symmetry. With this unit and timelike vector, the general action of the Einstein-{\AE}ther theory is given by \cite{jacobson_einstein-aether_2008}
	\bqn
	S_{\text{\ae}} = \frac{1}{16 \pi G_{\text{\ae}}} \int d^4 x \sqrt{-g} \Big(R+ \mathcal{L}_{\text{\ae}}\Big),
	\eqn
	where $g$ is the determinant of the four dimensional metric $g_{\mu\nu}$ of the space-time with the signatures $(-, +, +, +)$, $R$ is the Ricci scalar, $G_{\text{\ae}}$ is the the aether gravitational constant, and the Lagrangian of the \ae ther field $\mathcal{L}_{\text{\ae}}$ is given by
	\bqn
	\mathcal{L}_{\text{\ae}} \equiv -M^{\alpha \beta}_{\;\; \;\; \mu\nu} (D_\alpha w^\mu) (D_{\beta} w^\nu) + \lambda (g_{\alpha \beta} w^\mu w^\nu +1).\nb\\
	\eqn
	Here $D_{\alpha}$ denotes the covariant derivative with respect to $g_{\mu\nu}$, $\lambda$ is a Lagrangian multiplier, which guarantees that the aether four-velocity $w^{\alpha}$ is always timelike, and $M^{\alpha \beta}_{\;\; \;\; \mu\nu}$ is defined as \footnote{The parameters $(c_1, c_2, c_3, c_4)$ used in this paper are related to parameters $(c_\theta, c_\sigma, c_\omega, c_a)$ by the relations $c_{\theta} = c_1+3c_2+c_3$, $c_{\sigma } = c_1+c_3 = c_{13}$, $c_\omega = c_1- c_3$, $c_a = c_1+c_4 = c_{14}$.}
	\bqn
	M^{\alpha \beta}_{\;\; \;\; \mu\nu} \equiv c_1 g^{\alpha \beta} g_{\mu\nu} + c_2 \delta^{\alpha}_{\mu} \delta ^{\beta}_{\nu} + c_3  \delta^{\alpha}_{\nu} \delta ^{\beta}_{\mu} - c_4 w^{\alpha} w^{\beta} g_{\mu\nu}. \nb\\
	\eqn
	The four coupling constants $c_i$'s are all dimensionless, and $G_{\text{\ae}}$ is related to the Newtonian constant $G_N$ via the relation \cite{carroll_lorentz-violating_2004},
	\bqn
	G_{\text{\ae}} = \frac{G_N}{1-\frac{1}{2}c_{14}}
	\eqn
	with $c_{14} \equiv c_1+c_4$. In order to discuss the black hole solution with electric charge, we also add a source-free Maxwell Lagrangian $\mathcal{L}_{M}$ to the theory, then the total action of the theory becomes,
	\bqn
	S_{\text{\ae}, M} =  S_{\text{\ae}} + \int d^4x \sqrt{-g } \mathcal{L}_M,
	\eqn
	where
	\bqn
	\mathcal{L}_{M} = - \frac{1}{16 \pi G_{\text{\ae}}} F^{\mu\nu}F_{\mu\nu},\\
	F_{\mu\nu} = \nabla_\mu A_\nu - \nabla_\nu A_\mu,
	\eqn
	where $A_\mu$ is the electromagnetic four vector potential.
	
	The variations of the total action with respect to $G_{\mu\nu}$, $w^{\alpha}$, $\lambda$, and $A^a$ yield, respectively, the field equations,
	\bqn
	E^{\mu\nu} =0, \lb{einstein_equation}\\
	{\text{\AE}}_{\alpha} =0, \lb{aether_equation}\\
	g_{\mu\nu} w^{\mu} w^{\nu} =-1 \lb{lambda_equation},\\
	\nabla^\mu F_{\mu\nu}=0.
	\eqn
	where
	\bqn
	E^{\mu\nu} \equiv R^{\mu\nu} - \frac{1}{2} g^{\mu\nu} R - 8 \pi G_{\text{\ae}} T_{\text{\ae}}^{\mu\nu},\\
	\AE_{\alpha} \equiv D_{\mu} J^{\mu}_{\;\;\;\alpha} + c_4 a_{\mu} D_{\alpha}w^{\mu} + \lambda w_{\alpha},
	\eqn
	with
	\bqn
	T^{\text{\ae}}_{\alpha\beta} &\equiv& D_{\mu}\Big[J^{\mu}_{\;\;\;(\alpha}w_{\beta)} + J_{(\alpha\beta)}w^{\mu}-w_{(\beta}J_{\alpha)}^{\;\;\;\mu}\Big]\nb\\
	&& + c_1\Big[\left(D_{\alpha}w_{\mu}\right)\left(D_{\beta}w^{\mu}\right) - \left(D_{\mu}w_{\alpha}\right)\left(D^{\mu}w_{\beta}\right)\Big]\nb\\
	&& + c_4 a_{\alpha}a_{\beta}    + \lambda  w_{\alpha}w_{\beta} - \frac{1}{2}  g_{\alpha\beta} J^{\delta}_{\;\;\sigma} D_{\delta}w^{\sigma},\nb\\
	J^{\alpha}_{\;\;\;\mu} &\equiv& M^{\alpha\beta}_{~~~~\mu\nu}D_{\beta}w^{\nu},\nb\\
	a^{\mu} &\equiv& w^{\alpha}D_{\alpha}w^{\mu}.
	\eqn
	From Eqs.(\ref{aether_equation}) and (\ref{lambda_equation}),  we find that
	\bqn
	\lb{2.7}
	\lambda = w_{\beta}D_{\alpha}J^{\alpha\beta} + c_4 a^2,
	\eqn
	where $a^{2}\equiv a_{\lambda}a^{\lambda}$.
	
	Recently, as mentioned above,  the combination of the gravitational wave event GW170817 \cite{abbott_gw170817_2017}, observed by the LIGO/Virgo collaboration, and the event of the gamma-ray burst GRB170817A \cite{abbott_gravitational_2017} provides  a remarkably stringent constraint on the speed of the spin-2 mode, $- 3\times 10^{-15} < c_T -1 < 7\times 10^{-16}$. In the Einstein-{\AE}ther theory, the speed of the spin-2 graviton is given by $c_{T}^2 = 1/(1-c_{13})$ \cite{jacobson_einstein-eather_2004} with $c_{13} \equiv c_1+c_3$, so the  GW170817 and GRB 170817A events imply
	\bq
	\lb{2.8a}
	\left |c_{13}\right| < 10^{-15}.
	\eq
	Together with other observational and theoretical constraints, recently it was found that the parameter space of the theory is further restricted to \cite{oost_constraints_2018}
	\bq
	\lb{2.8b}
	c_{4} \lesssim 0, \quad 0 \lesssim c_2 \lesssim 0.095, \quad 0 \lesssim c_{14} \lesssim 2.5\times 10^{-5}.
	\eq
	
	\subsection{Static and Charged Spherically Symmetric Einstein-{\AE}ther Black Holes}
	
	The general form for a static spherically symmetric metric for Einstein-\ae ther black hole spacetimes can be written in the Eddington-Finklestein coordinate system as
	\bqn \lb{metric_EF}
	ds^2 = -e(r) dv^2 +2 f(r) dv dr + r^2 \gamma_{ij} dx^i dx^j,
	\eqn
	with the corresponding killing vector $\chi^a$ and the \ae ther vector field $w^{a}$ being given by
	\bqn
	\chi^a=(1,0,0,0),\;\;\; w^{a} = (\alpha, \beta, 0, 0 ),
	\eqn
	where $e(r)$, $f(r)$, $\alpha(r)$, and $\beta(r)$ are functions of $r$ only, and $\gamma_{ij}$ represents the metric of the two-dimensional sphere $S^2$. The boundary conditions on the metric components are such that the solution is asymptotically flat, while those for the aether components are such that
	\bqn
	\lim_{r \to +\infty} w^a = (1,0,0,0).
	\eqn
	
	As shown in \cite{ding_charged_2015}, there exist two types of exact static and charged spherically symmetric black hole solutions in Einstein-\ae ther theory. The first solution corresponds to the special choice of coupling constants $c_{14}=0$ and $c_{123} \neq 0$ where $c_{123} \equiv c_1+c_2+c_3$, while the second solution corresponds to $c_{123}=0$.
	
	\subsubsection{\texorpdfstring{$c_{14}=0$}{c14=0} and \texorpdfstring{$c_{123} \neq 0$}{c123=/=0}}
	
	For the first solution, we have \cite{ding_charged_2015}
	\bqn
	e(r) &=& 1- \frac{2GM}{c^2r} +\frac{GQ^2}{4\pi\epsilon_0 c^4 r^2} \nb\label{e1st}\\
	&&- \frac{27 c_{13}}{256(1-c_{13})} \left(\frac{2GM}{c^2r} \right)^4,\lb{e14}\\
	f(r) &=& 1,\\
	\alpha(r) &=& \Bigg[ \frac{1}{\sqrt{1-c_{13}}} \frac{3\sqrt{3}}{16} \left(\frac{2M}{r}\right)^2 \nb\\
	&&~~~  + \sqrt{1-\frac{2M}{r} + \frac{27}{256} \left(\frac{2M}{r}\right)^4}\Bigg]^{-1},\\
	\beta(r) &=& - \frac{1}{\sqrt{1-c_{13}}} \frac{3 \sqrt{3}}{16} \left(\frac{2M}{r}\right)^2.
	\eqn
	Here $M$ and $Q$ are the mass and the electric charge of the black hole spacetime respectively. It is obvious that when $c_{13}=0$, the above solution reduces to the Reissner--Nordstr\"{o}m black hole.

	\subsubsection{\texorpdfstring{$c_{123}=0$}{c123=0}}
	
	For the second solution, we have \cite{ding_charged_2015}
	\bqn
	e(r) &=& 1- \frac{2GM}{c^2r}  +\frac{1}{1-c_{13}}~\frac{GQ^2}{4\pi\epsilon_0 c^4 r^2} \nb \label{e2nd}\\
	&&-\frac{2c_{13} - c_{14}}{8(1-c_{13})} \left(\frac{2GM}{c^2r} \right)^2, \lb{e123}\\
	f(r) &=& 1,\\
	\alpha(r) &=&  \frac{1}{1+\frac{1}{2} \left[\sqrt{\frac{2-c_{14}}{2(1-c_{13})}} -1 \right]\frac{2M}{r}},\\
	\beta(r) &=& - \frac{1}{2} \sqrt{\frac{2-c_{14}}{2(1-c_{13})}} \frac{2M}{r} .
	\eqn
	In this case, when $c_{13}=0=c_{14}$, it also reduces to the Reissner--Nordstr\"{o}m black hole.
	
	For both solutions~\eqref{e1st} and~\eqref{e2nd} we have introduced some physical constants only in the two expressions of $e(r)$. These constants, along with the solar mass $M_\odot$, are needed in Sec.~\ref{subsecres}. In SI units they assume the numerical values $M_\odot=1.9888\times 10^{30}$, $G=6.673\times 10^{-11}$ and $c=299792458$. These same constants will be written explicitly in some subsequent formulas that are used in Sec.~\ref{subsecres}.
	
	For both solutions, it is convenient to write the metric (\ref{metric_EF}) in the the Eddington-Finklestein coordinate system in the form of the usual $(t, \; r,\; \theta, \; \varphi)$ coordinates. This can be achieved by using the coordinate transformation
	\bqn
	dt = dv - \frac{dr }{e(r)}, \;\; dr=dr.
	\eqn
	Then the metric of the background spacetime is in the form
	\bqn\lb{metric}
	ds^2 = - c^2 e(r)dt^2+\frac{dr^2}{e(r)} + r^2 (d\theta^2
	+\sin^2\theta d\varphi ^2).
	\eqn
	In this metric, the \ae{}ther field reads
	\bqn
	w^a= \left(\alpha(r) - \frac{\beta(r)}{e(r)}, \beta(r), 0, 0\right).
	\eqn

	\section{Quasi-periodic oscillations (QPOs)\label{sececos}}
	Motion perturbations around stable timelike paths yields epicyclic oscillations, often called the quasi-periodic oscillations, the frequencies of which have direct observational effects~\cite{qpos1}-\cite{res2}.
	
	The most concerned of all stable timelike paths are circular orbits in the plane of symmetry, which are the trajectories borrowed by in-falling matter in accretion processes. In this section we will derive the equations governing motion perturbations around stable circular orbits in their general form. Some of the equations derived in this section will be used in some parts of Sec.~\ref{secgeod} dealing with the geodesic motion.

	From now on we consider stable paths in the $\theta=\pi/2$ plane. Since the metric element $e(r)$ does not depend on $\theta$ one may show~\cite{qposknb} that the only nonvanishing elements of $\Gamma^{\mu}_{\alpha\beta}$, $\partial_r\Gamma^{\mu}_{\alpha\beta}$, and $\partial_{\theta}\Gamma^{\mu}_{\alpha\beta}$ \emph{on the $\theta=\pi/2$ plane} are
	\begin{align}
	\label{Ga}&\Gamma^{t}_{tr},\;\Gamma^{r}_{tt},\;\Gamma^{r}_{rr},
	\;\Gamma^{r}_{\theta\theta},\;\Gamma^{r}_{\varphi\varphi},\;\Gamma^{\theta}_{r\theta},\;\Gamma^{\varphi}_{r\varphi},\\
	\label{Gb}&\partial_r\Gamma^{t}_{tr},\;\partial_r\Gamma^{r}_{tt},\;\partial_r\Gamma^{r}_{rr},
	\;\partial_r\Gamma^{r}_{\theta\theta},\;\partial_r\Gamma^{r}_{\varphi\varphi},\;\partial_r\Gamma^{\theta}_{r\theta},\;\partial_r\Gamma^{\varphi}_{r\varphi},\\
	\label{Gc}&\partial_{\theta}\Gamma^{\theta}_{\varphi\varphi},\;\partial_{\theta}\Gamma^{\varphi}_{\theta\varphi},
	\end{align}
	and those obtained by symmetry.

	\subsection{Unperturbed and perturbed circular motions\label{pcm}}
	Any small deviation from a \emph{stable} circular motion leads to an epicyclic motion around the stable circular path. There are two types of deviations: A radial deviation in the $\theta=\pi/2$ plane and a vertical deviation perpendicular to the $\theta=\pi/2$ plane. One can perform both deviations at the same time.
	
	First of all, we need to set the equations governing an unperturbed circular motion. Once this is done, we will derive the equations that describe a perturbed circular motion around a stable unperturbed circular motion. In a third step we will work to decouple the set of equations governing the perturbed circular motion.

	\subsubsection{Unperturbed circular motion}
	Unperturbed circular motion is a geodesic motion obeying the equation,
	\begin{equation}\label{p1}
	\frac{d u^{\mu}}{d \tau}+\Gamma^{\mu}_{\alpha\beta}u^{\alpha}u^{\beta}=
	0,
	\end{equation}
	where $u^{\mu}=d x^{\mu}/d \tau =\dot x^{\mu}$ is the four-velocity. Here the connection $\Gamma^{\mu}_{\alpha\beta}$ is related to the unperturbed metric~\eqref{metric}.
	
	For a circular motion in the equatorial plane ($\theta=\pi/2$), $u^{\mu}=(u^t,\,0,\,0,\,u^{\varphi})=u^t(1,\,0,\,0,\,\omega)$, where $\omega=d\varphi/d t$ is the angular velocity. The only equations describing such a motion are the $r$ component of~\eqref{p1} and the normalization condition $g_{\mu\nu}u^{\mu}u^{\nu}=-c^2$, which take the following forms~\cite{qposknb}, respectively
	\begin{align}
	\label{p2a}&\partial _r g_{tt} (u^t)^2+\partial _r g_{\varphi \varphi } (u^{\varphi })^2=0,\\
	\label{p2b}&g_{tt} (u^t)^2+g_{\varphi \varphi } (u^{\varphi })^2=-c^2,
	\end{align}
	where $g_{tt}=-c^2 e$~\eqref{metric} and the metric and its derivatives are all evaluated at $\theta=\pi/2$. These equations can be solved for ($u^t,\,u^{\varphi }$) in terms of the radius $r_0$ of the circle
	\begin{align}
	\label{p3a}&u^t(r_0)=c\sqrt{\frac{\partial _r g_{\varphi \varphi }}{g_{\varphi \varphi }\partial _r g_{tt}-g_{tt} \partial
			_r g_{\varphi \varphi }}}=\sqrt{\frac{2}{2e(r_0)-r_0e'(r_0)}}\,,\\
	\label{p3b}&u^{\varphi }(r_0)=\pm c\sqrt{\frac{-\partial _r g_{tt}}{g_{\varphi \varphi }\partial _r g_{tt}-g_{tt} \partial
			_r g_{\varphi \varphi }}}\nonumber\\
	&\qquad	  =\pm c r_0^{-1/2}\sqrt{\frac{e'(r_0)}{2e(r_0)-r_0e'(r_0)}}\,,
	\end{align}
	where all function are evaluated at $r_0$ and $\theta=\pi/2$.

	\subsubsection{Perturbed circular motion: ECOs\label{ECOs}}
	If the motion is perturbed, the actual position is now denoted by $X^{\mu}=x^{\mu}+\eta^{\mu}$ and the 4-velocity by $U^{\mu}=u^{\mu}+\dot{\eta}^{\mu}$ (where $~\dot{}\equiv d /d\tau$) with $u^{\mu}$ are the unperturbed values given by~\eqref{p3a} and~\eqref{p3b}. Using this in
	\begin{equation}
	\frac{d U^{\mu}}{d \tau}+\Gamma^{\mu}_{\alpha\beta}(X^{\sigma})U^{\alpha}U^{\beta}=
	0,
	\end{equation}
	along with~\eqref{p1} and keeping only linear terms in $\eta^\mu$ and its derivatives we arrive at
	\begin{equation}\label{p4}
	\ddot{\eta}^{\mu}+2\Gamma^{\mu}_{\alpha\beta}u^{\alpha}\dot{\eta}^{\beta}
	+\partial_{\nu}\Gamma^{\mu}_{\alpha\beta}u^{\alpha}u^{\beta}\eta^{\nu}
	=0,
	\end{equation}
	where $\Gamma^{\mu}_{\alpha\beta}$ and its derivatives are evaluated at $\theta=\pi/2$. This relation was also derived in~\cite{Kerr1}. \\

	\paragraph{The $\theta$ component.}
	Using~\eqref{Ga} to~\eqref{Gc} in~\eqref{p4} the $\theta$ component decouples and takes the form of an oscillating vertical motion (perpendicular to the $\theta=\pi/2$ plane)~\cite{qposknb}
	\begin{align}
	\label{p5a}&\ddot{\eta}^{\theta}+\Omega_{\theta}^2\eta^{\theta}=0,\\
	&\Omega_{\theta}^2(r_0)\equiv \partial_{\theta}\Gamma^{\theta}_{\varphi\varphi}(u^{\varphi})^2=(u^{\varphi})^2=\frac{c^2e'(r_0)}{r_0[2e(r_0)-r_0e'(r_0)]},
	\end{align}
	where the derivative with respect to $\theta$ is evaluated at $\theta=\pi/2$. Since $u^{\varphi}$ is function of the radius of the circle $r_0$, $\Omega_{\theta}^2$ depends on $r_0$ too. We see that the epicyclic frequency $\Omega_{\theta}$ is just the Keplerian or orbital frequency in absolute value. The stability of the circular motion is partially guaranteed by the positiveness of $\Omega_{\theta}^2$. In the regions of space where $u^{\varphi}$ is real, the vertical motion is always stable. \\

	\paragraph{The $t$ and $\varphi$ components.}
	Expressions for $\dot{\eta}^{t}$ and $\dot{\eta}^{\varphi}$ are obtained from the $t$ and $\varphi$ components of~\eqref{p4} by integration
	\begin{equation}\label{p6}
	\dot{\eta}^{j}=-2\Gamma^{j}_{ri}u^i\eta^r,\qquad (i,\,j=t,\,\varphi),
	\end{equation}
	where we have set the constants of integration to zero.\\

	\paragraph{The $r$ component.}
	Using~\eqref{Ga} to~\eqref{Gc} the $r$ equation of~\eqref{p4} is first brought to the form
	\begin{multline}\label{p7}
	\ddot{\eta}^{r}+\partial_{r}\Gamma^{r}_{ij}u^iu^j\eta^{r}
	+2\Gamma^{r}_{ji}u^i\dot{\eta}^{j}=0,
	\qquad (i,\,j=t,\,\varphi),
	\end{multline}
	then, using~\eqref{p6} to eliminate $\dot{\eta}^{j}$ ($j=t,\,\varphi$), we decouple it into an equation describing an oscillating radial motion,
	\begin{equation}\label{p8a}
	\ddot{\eta}^{r}+\Omega_{r}^2\eta^{r}=0,
	\end{equation}
	with frequency $\Omega_{r}$ given by~\cite{qposknb}
	\begin{align}\label{p8b}
	&\Omega_{r}^2(r_0)\equiv (\partial_{r}\Gamma^{r}_{ij}-4\Gamma^{r}_{ik}\Gamma^{k}_{rj})u^iu^j,\quad (i,\,j,\,k=t,\,\varphi),\nonumber\\
	&\qquad\quad = \frac{c^2}{2} (u^t)^2 (e e''-e'^2)+(u^{\varphi })^2 (3 e-r_0 e'),\nonumber\\
	&\qquad\quad =c^2~\frac{3 e(r_0) e'(r_0)-2 r_0 e'(r_0)^2+r_0 e(r_0) e''(r_0)}{r_0[2 e(r_0)-r_0 e'(r_0)]},
	\end{align}
	where the function $e$ and its derivatives are evaluated at $r_0$. We see that $\Omega_{r}$ is nonlinearly coupled to the orbital frequency $u^\varphi$. The stability of the oscillating radial motion is ensured by the positiveness of $\Omega_{r}^2$. The circular motion is considered stable if both local frequencies $\Omega_{r}^2$ and $\Omega_{\theta}^2$ are positive.
	
	Using the definition of ($E,\,L$): $g_{tt} u^{t} =-c^2 e u^{t}= - E$ and $g_{\varphi \varphi} u^{\varphi } = L$ [see Eq.~\eqref{phidot}] along with~\eqref{p3a} and~\eqref{p3b} we determine ($E,\,L$) in terms of $r_0$
	\begin{multline}\label{EL}
	E=c^2e(r_0)\sqrt{\frac{2}{2e(r_0)-r_0e'(r_0)}}\,,\\ L=\pm cr_0^{3/2}\sqrt{\frac{e'(r_0)}{2e(r_0)-r_0e'(r_0)}}\,.
	\end{multline}
	\begin{figure*}[!htb]
		\centering
		\includegraphics[width=0.45\textwidth]{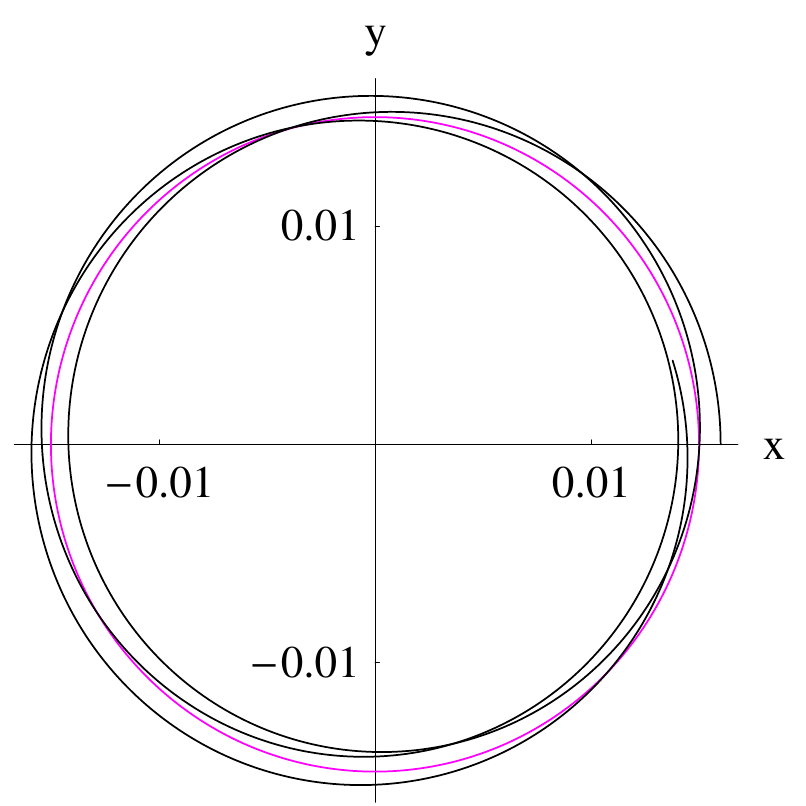}
		\includegraphics[width=0.45\textwidth]{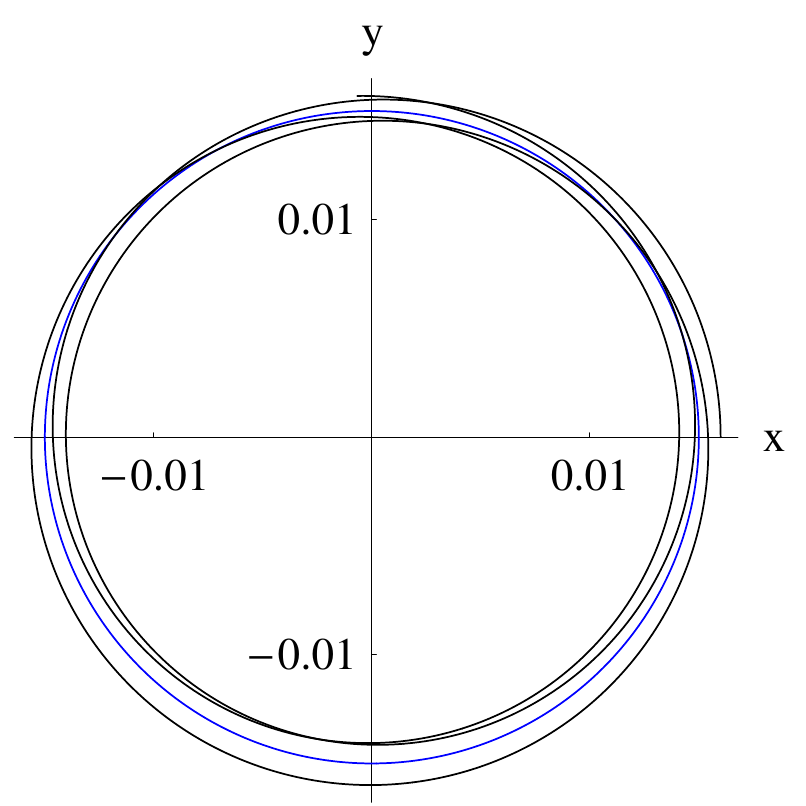}
		\caption{Stability of the radially perturbed motion around the circle $r_0=0.015$ for $M=1/1000$ and $Q=0$. The motion is in the trigonometric direction. The perturbed motion started off at the rightmost point $r=0.016$, $\varphi=0$. Left Plot: The Schwarzschild black hole. The black plot oscillates around the magenta circle $r_0=0.015$ with a frequency $\Omega_{r}(r_0)=14.9071$. The proper time laps between two successive encounters of the black and magenta plots corresponds to a half period of the oscillatory motion. The first encounter happens in the second quadrant and the second one in the fourth quadrant (almost near the $x$-axis). For this set of parameters the event horizon is at $r_\text{h}=1/500=0.002$ and $r_\text{isco}=2r_{\text{ph}}=3/500=0.006$. Right Plot: The first solution~\eqref{e14} for $c_{13}=0.99$. The black plot oscillates around the blue circle $r_0=0.015$ with a frequency $\Omega_{r}(r_0)=11.7848$. The proper time laps between two successive encounters of the black and blue plots corresponds to a half period of the oscillatory motion. The first encounter happens in the second quadrant and the second one in the first quadrant. For this set of parameters the event horizon is at $r_\text{h}=0.0022$, $r_\text{isco}=0.0118>2r_{\text{ph}}$ and $r_{\theta\text{-isco}}=0.0057$.}
		\label{figqpos01}
	\end{figure*}
	\begin{figure}[!htb]
		\centering
		\includegraphics[width=0.49\textwidth]{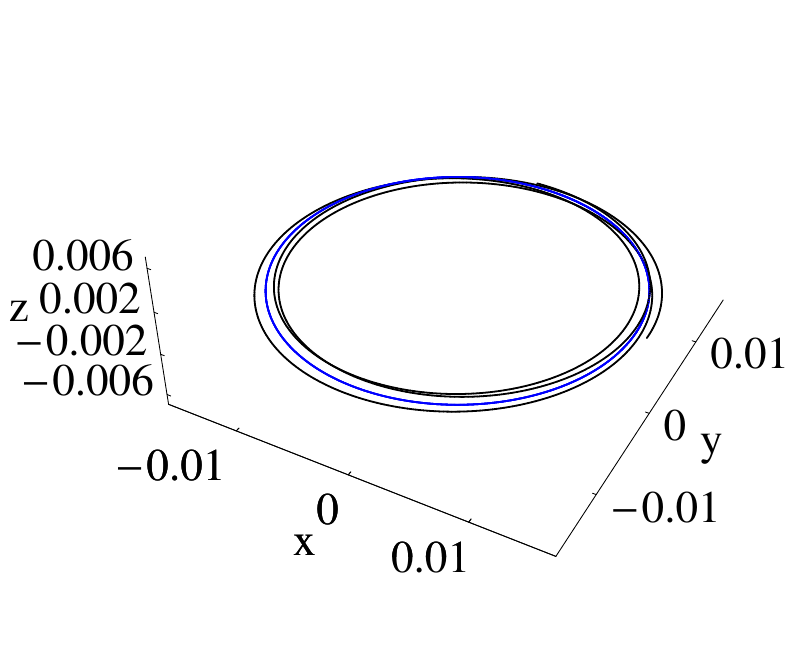}
		\caption{3D diagram. Stability of the radially and vertically perturbed motion around the blue circle $r_0=0.015>r_\text{isco}$ and $\theta=\pi/2$ for the first solution~\eqref{e14} taking $M=1/1000$, $Q=0$ and  $c_{13}=0.99$. Here $r_\text{isco}=0.0118$. The toroidal motion of the particle is in the trigonometric direction as seen from above. The perturbed motion started off at the point $r=0.016$, $\theta=\pi/2-0.0332$, $\varphi=0$. The black plot oscillates around the blue circle with a radial frequency $\Omega_{r}(r_0)=11.7848$ and a vertical frequency $\Omega_{\theta}(r_0)=20.3012$.}
		\label{figqpos3d}
	\end{figure}
	\begin{figure}[!htb]
		\centering
		\includegraphics[width=0.49\textwidth]{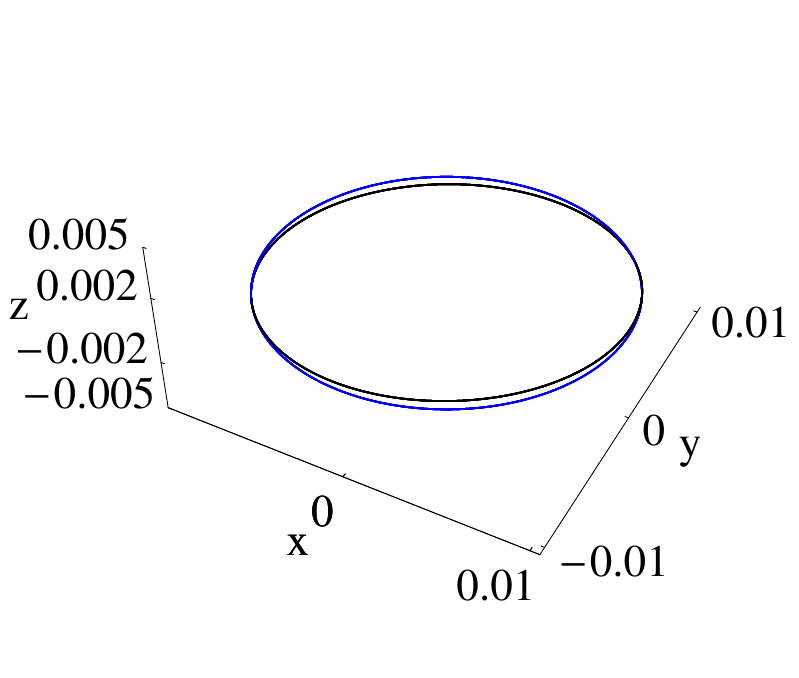}
		\caption{3D diagram. Stability of the vertically perturbed motion around the blue circle $r_{\theta\text{-isco}}<r_0=0.01<r_\text{isco}$ and $\theta=\pi/2$ for the first solution~\eqref{e14} taking $M=1/1000$, $Q=0$ and  $c_{13}=0.99$. Here $r_\text{isco}=0.0118>2r_{\theta\text{-isco}}$ and $r_{\theta\text{-isco}}=0.0057$. The motion is in the trigonometric direction as seen from above. The perturbed motion started off at the point $r=0.01$, $\theta=\pi/2-0.0332$, $\varphi=0$. The black plot oscillates around the blue circle with a vertical frequency $\Omega_{\theta}(r_0)=45.3087$. Since there are oscillations in the region enclosed by $r_\text{ph}$ and $r_\text{isco}$, the energy radiated there should not be neglected.}
		\label{figqpos3dt}
	\end{figure}
	\begin{figure*}[!htb]
		\centering
		\includegraphics[width=0.33\textwidth]{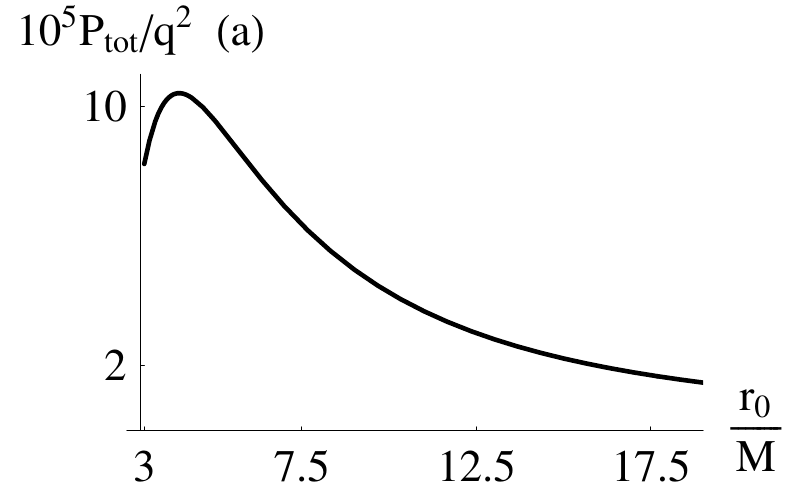}
		\includegraphics[width=0.33\textwidth]{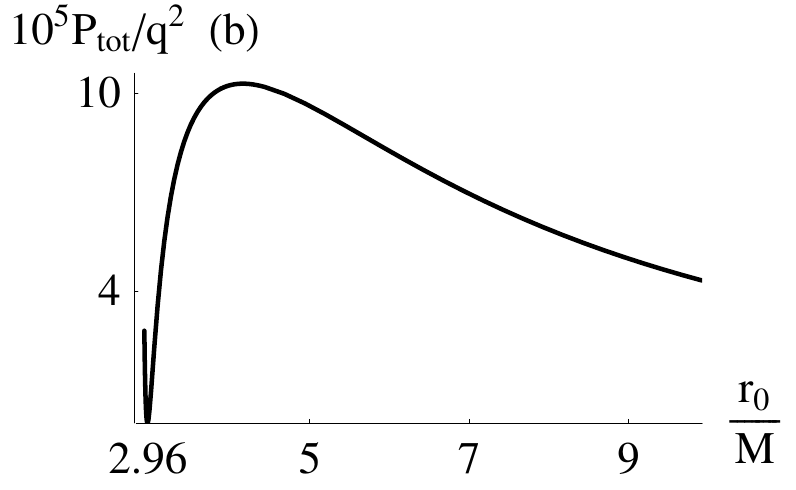}
		\includegraphics[width=0.33\textwidth]{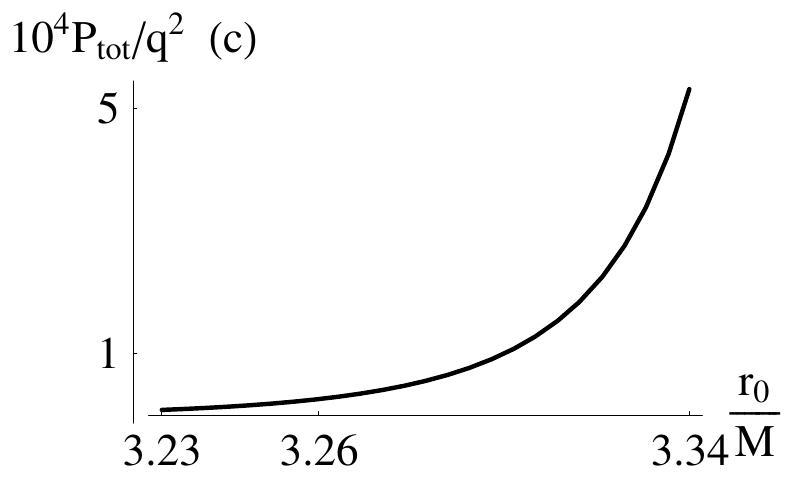}
		\caption{The electromagnetic radiation intensity per unit charge squared $P_\text{tot}/q^2$ versus the radius $r_0/M$ of the circle for the first solution~\eqref{e14} taking $Q=0$. In these plots the minimum value of $r_0$ corresponds to $2e(r_0)-r_0e'(r_0)=0$ ($\Omega_\theta^2=\infty$): This is the radius of the circular photon orbit denoted by $r_{\text{ph}}$. Plot (a): Schwarzschild metric ($c_{13}=0$ and $r_{\text{ph}}/M=3$). Plot (b): Metric~\eqref{e14} with $c_{13}=-0.6$ ($r_{\text{ph}}/M=2.92$). The value $r_0/M=2.96$ is where $P_\text{tot}$ vanishes. Plot (c): Metric~\eqref{e14} with $c_{13}=0.6$ ($r_{\text{ph}}/M=3.23$). The corresponding graphs for the second solution~\eqref{e2nd} are similar to these ones and they depend on the values of the parameters $c_{13}$ and $c_{14}$ as described in the last paragraph of Sec.~\ref{subsecrad}.}
		\label{figrad}
	\end{figure*}
	\begin{figure*}[!htb]
		\centering
		\includegraphics[width=0.33\textwidth]{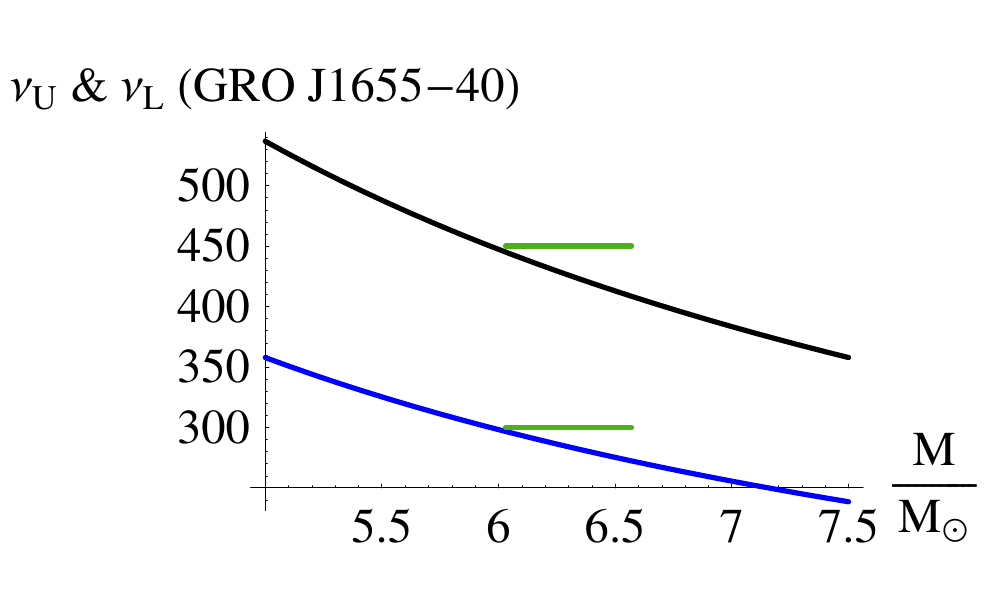}
		\includegraphics[width=0.33\textwidth]{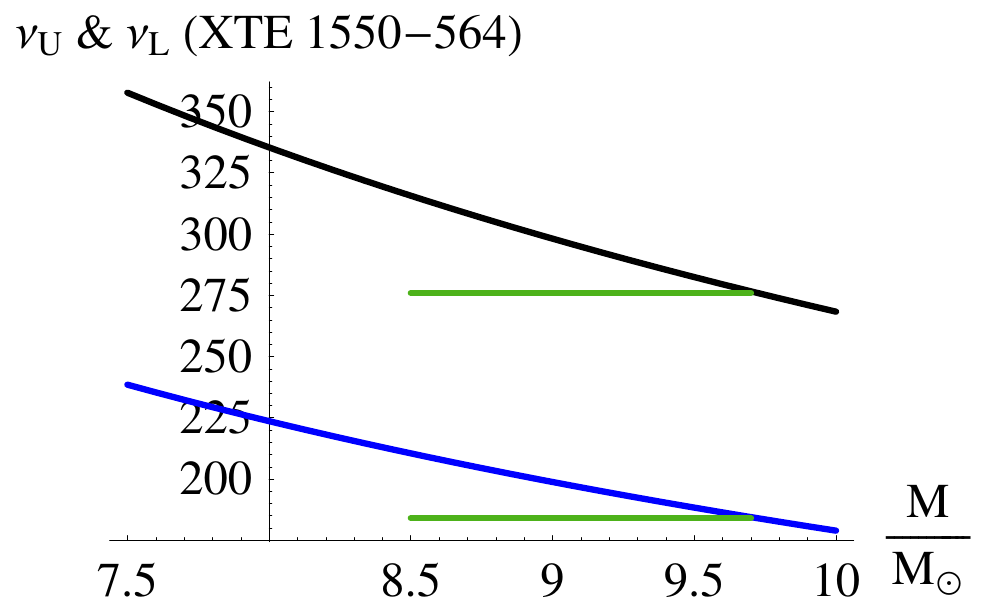}
		\includegraphics[width=0.33\textwidth]{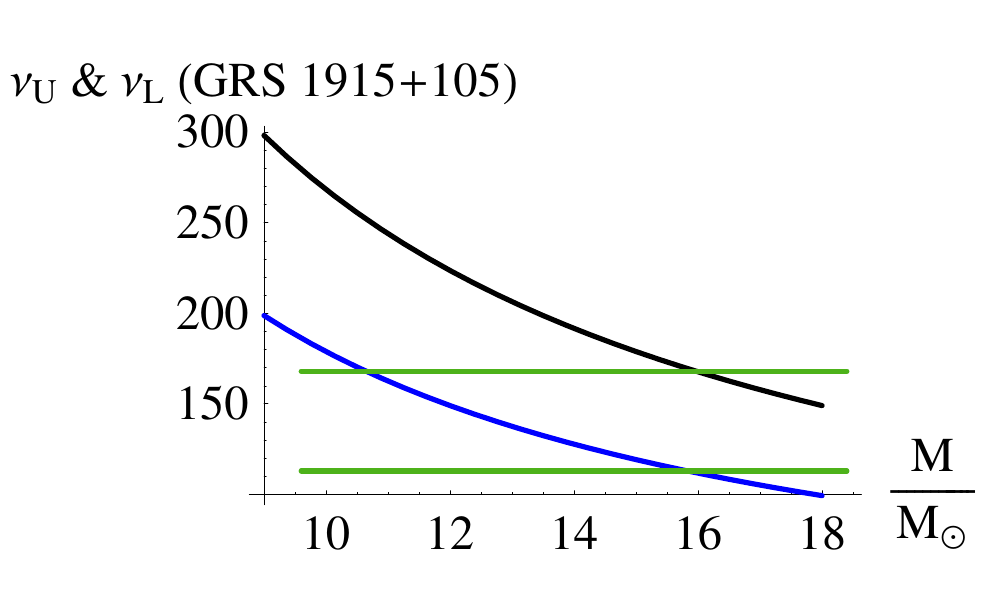}
		\caption{Fitting the uncharged particle oscillation upper and lower frequencies to the observed frequencies in Hz for the microquasars GRO J1655-40, XTE J1550-564 and GRS 1915+105 at the 3/2 resonance radius for the first solution~\eqref{e14} taking $Q=0$ and $c_{13}=-0.6$. The black plot represents $\nu_U$, the blue plot represents $\nu_L$ and the green lines represent the mass limits as given in~\eqref{pr1}, \eqref{pr2} and~\eqref{pr3}. Each curve (black or blue) crosses the (upper or lower) mass error band of each microquasar for the same value of $c_{13}=-0.6$.}
		\label{figres}
	\end{figure*}
	\begin{figure*}[!htb]
		\centering
		\includegraphics[width=0.33\textwidth]{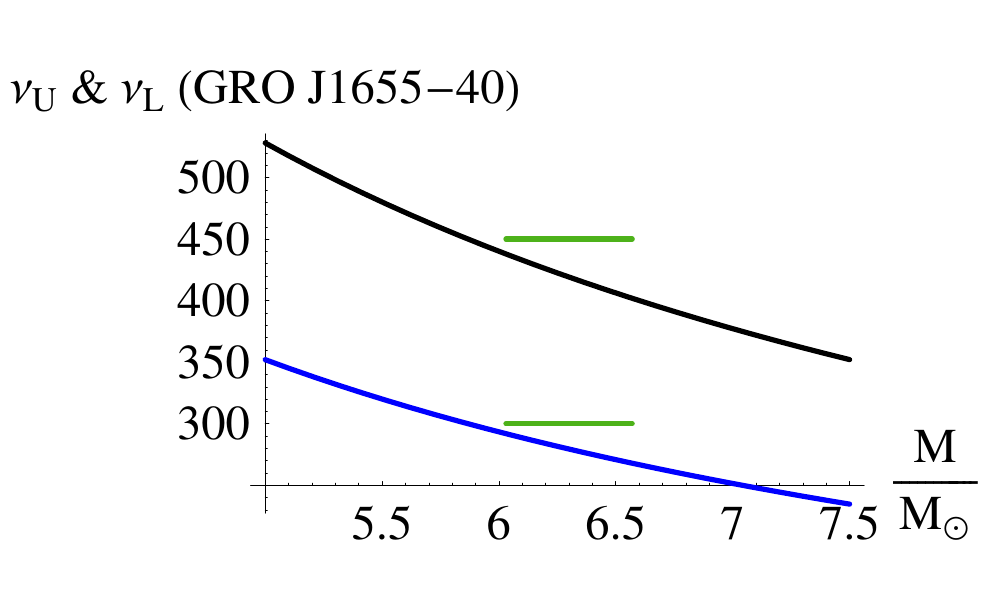}
		\includegraphics[width=0.33\textwidth]{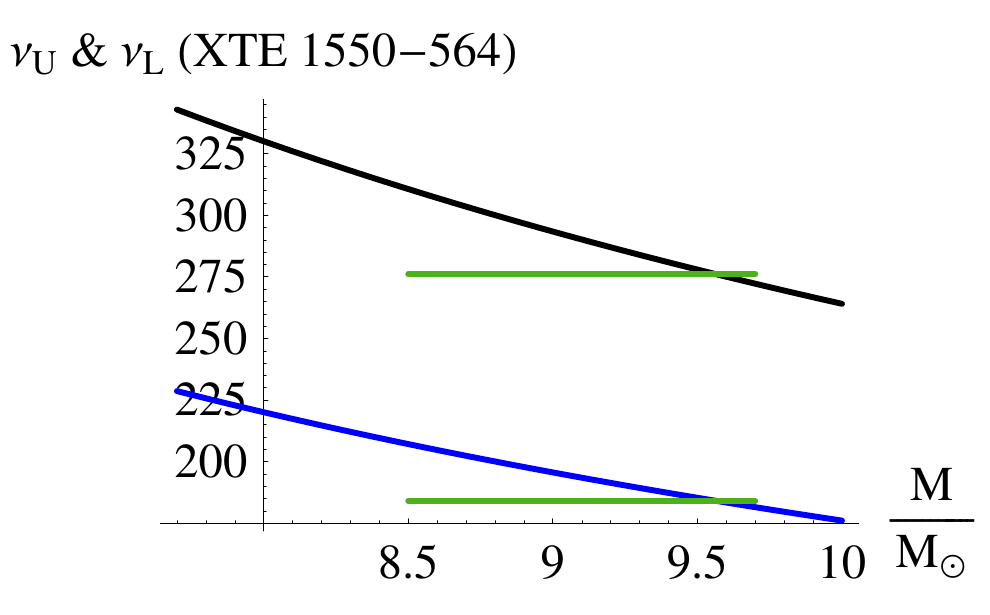}
		\includegraphics[width=0.33\textwidth]{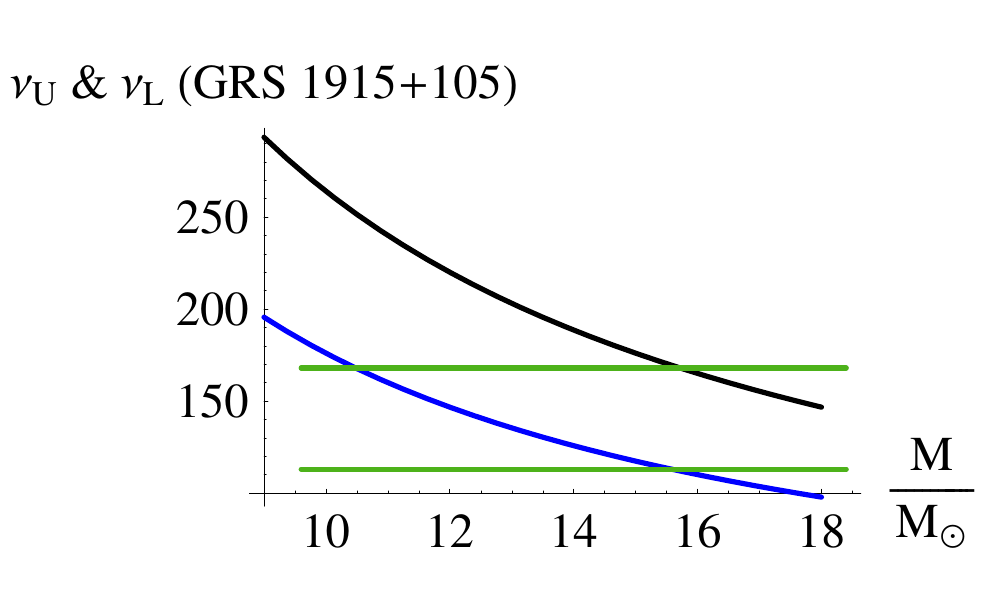}
		\caption{Fitting the uncharged particle oscillation upper and lower frequencies to the observed frequencies in Hz for the microquasars GRO J1655-40, XTE J1550-564 and GRS 1915+105 at the 3/2 resonance radius for the Schwarzschild metric ($c_{13}=0$). The black plot represents $\nu_U$, the blue plot represents $\nu_L$ and the green lines represent the mass limits as given in~\eqref{pr1}, \eqref{pr2} and~\eqref{pr3}. Only for the microquasar GRO J1655-40 the black and blue cures do not cross the mass error bands but they pass very adjacent to them.}
		\label{figresS}
	\end{figure*}
	\begin{figure}[!htb]
	\centering
	\includegraphics[width=0.42\textwidth]{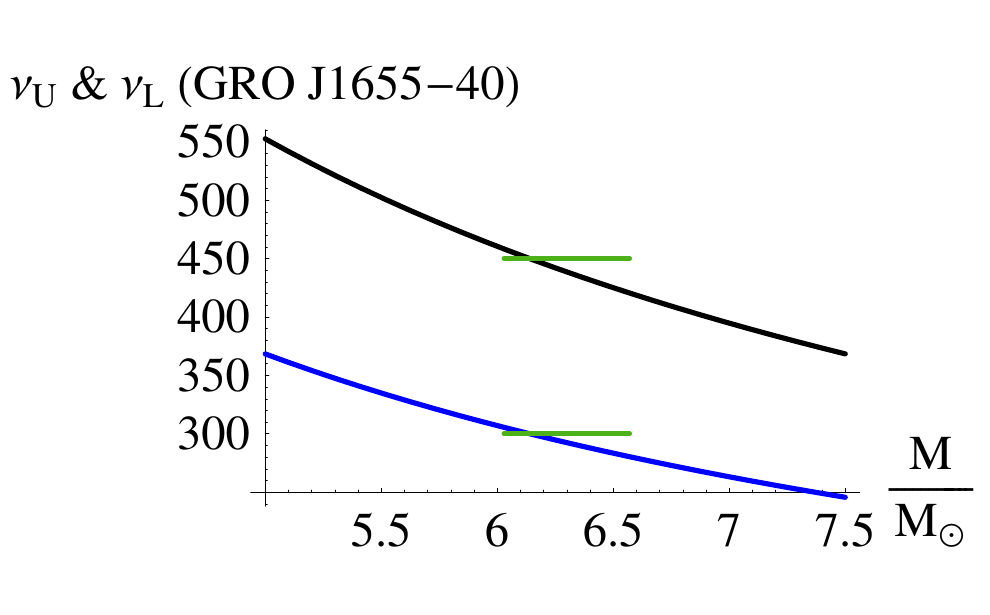}
	\caption{Fitting the uncharged particle oscillation upper and lower frequencies to the observed frequencies in Hz for the microquasar GRO J1655-40 at the 3/2 resonance radius for the first solution~\eqref{e14} taking $Q=0$ and $c_{13}=-5000$. The black plot represents $\nu_U$, the blue plot represents $\nu_L$ and the green lines represent the mass limits as given in~\eqref{pr1}, \eqref{pr2} and~\eqref{pr3}. Each curve (black or blue) crosses the (upper or lower) mass error band of the microquasar.}
	\label{figres5000}
	\end{figure}
	
	In summarizing we have relied on the properties~\eqref{Ga} to~\eqref{Gc} to obtain the decoupled equations for $\eta^r$ and $\eta^{\theta}$.
	
	Following~\cite{qposknb} we determine the relation of $\Omega_{r}$ to the effective potential $V_{\text{eff}}$ by
	\begin{equation}\label{f4}
	\Omega_{r}^2(r_0)=\frac{V_{\text{eff}}''(r_0)}{2},
	\end{equation}
	that is, in the representation $\dot r ^2 = E^2 - V_{\rm eff}(r)$ [see Eq.~\eqref{rdot}], the second derivative of the potential at $r_0$ should be positive to have a stable path there: $V_{\text{eff}}''(r_0)>0$. If the motion is circular, $r_0$ is a constant and $V_{\text{eff}}(r_0)=V_{\text{eff}}'(r_0)=0$, so the condition $V_{\text{eff}}''(r_0)>0$ implies that the potential should have a minimum value at $r_0$ to ensure stability of the circular orbit.
	
	At the inner stable circular orbit (isco) $V_{\text{eff}}''(r_{\text{isco}})=0$ and this consists a limiting case of stable circular orbits. This extra condition yields
	\begin{equation}\label{f5}
	\Omega_{r}^2(r_{\text{isco}})=0.
	\end{equation}
	Using~\eqref{p8b}, this reduces to
	\begin{multline}\label{r_isco}
	(3 e e'-2 r e'^2+r e e'')\Big|_{r=r_{\text{isco}}}=0\\ \Rightarrow \big(r^3e'e^{-2}\big)'\Big|_{r=r_{\text{isco}}}=0.
	\end{multline}
	
	From now on, we set the constraints $2e(r_0)>r_0e'(r_0)>0$
	ensuring that ($E,\,L$) are real numbers (different from 0) and that a circular motion is possible. This constraint ensures that all circular orbits are vertically stable against small oscillations. Along with the condition $\Omega_{r}^2(r_0)>0$, which ensures the radial stability against small oscillations, a circular motion will be vertically and radially stable if
	\begin{align}
	\label{cons1}& 2e(r_0)>r_0e'(r_0)>0,\quad\text{and}\\
	\label{const2}&3 e(r_0) e'(r_0)-2 r_0 e'(r_0)^2+r_0 e(r_0) e''(r_0)>0.
	\end{align}
	The constraint~\eqref{const2} is rewritten as $\big(r^3e'e^{-2}\big)'>0$ implying that the function $r^3e'e^{-2}$ should be increasing. For an asymptotically flat solution, this constraint is always satisfied for large $r$ since $e\simeq 1-2M/r$ yielding $r^3e'e^{-2}\simeq 2Mr$, which is manifestly increasing and approaches $\infty$ as $r\to\infty$. Near the horizon $r_\text{h}$ [$e(r_\text{h})=0$], the function $r^3e'e^{-2}$ approaches $\infty$ too as $r\to r_\text{h}^+$. So, the function $r^3e'e^{-2}$ must have at least one local extreme value. The nearest of such extrema to the horizon is the isco~\eqref{r_isco}.  
	
	We introduce the notion of $r_{\theta\text{-isco}}<r_{\text{isco}}$ to refer to the radius $r$ where the denominator of $\Omega_{\theta}$ vanishes: $\Omega_{\theta}(r_{\theta\text{-isco}})=\infty$. Circular orbits located in the region enclosed by $r_{\theta\text{-isco}}$ and $r_{\text{isco}}$ are only vertically stable since there $\Omega_{\theta}^2>0$ and $\Omega_{r}^2<0$. The radius $r_{\theta\text{-isco}}$ provides the innermost boundary (imb) of circular
	orbits for massive particles and it is the radius of the circular photon orbit denoted also by $r_{\text{ph}}$ or $r_{\text{imb}}$. In this work we will stick to the notation $r_{\text{ph}}$. For the Schwarzschild black hole we have $r_\text{isco}=2r_{\text{ph}}=6M$. As we shall see in Sec.~\ref{subsecapp}, for some choice of parameters, we have $r_\text{isco}>2r_{\text{ph}}$ for the first solution~\eqref{e14}.
	
	\subsection{Numerical considerations\label{subsecapp}}
	The small quantities $\eta^{\mu}$ are obtained by integrating the above-determined equations~\eqref{p5a}, \eqref{p6}, and~\eqref{p8a}
	\begin{align}\label{p9}
	&\eta^r=C_r\cos\Omega_r \tau,\quad \eta^{\theta}=C_{\theta}\sin[\Omega_{\theta} (\tau -\tau_0)],\nonumber\\
	&\eta^j=-2\Gamma^{j}_{ri}u^iC_r~\frac{\sin\Omega_r \tau}{\Omega_r},\qquad (i,\,j=t,\,\varphi),
	\end{align}
	where we have set the irrelevant constants to 0. The equation $\eta^t=-(2\Gamma^{t}_{ri}u^iC_r\sin\Omega_r \tau)/\Omega_r$ is not needed for the purpose of this section.
	
	In the following calculations we take $C_r=0.001$, $C_\theta=-0.05$, $\tau_0=0.5$ and $0\leq\tau\leq 1$. Since the peaks of the epicyclic frequencies are inversely proportional to $M$, it would be better to assign $M$ relatively smaller values to have smaller values of the periods $T_r\equiv 2\pi/\Omega_r$, $T_\theta\equiv 2\pi/\Omega_\theta$, allowing better graphical illustrations. Thus, in all graphs related to this section we took $M=1/1000$ and $Q=0$. We choose $r_0$ to be closer to the peak of $\Omega_r$: $r_0=0.015$. For this choice of parameters, we have $r_\text{isco}>2r_{\text{ph}}$ for the first solution~\eqref{e14} and no circular path is allowed for the second solution~\eqref{e2nd}.
	
	The results of calculations are summarized in Figs.~\ref{figqpos01} to~\ref{figqpos3dt}. In Fig.~\ref{figqpos01} we compare the stability of the radial perturbed circular motion in the Schwarzschild black hole, where $\Omega_{r}(r_0)=14.9071$, with that of the first solution~\eqref{e14}, where $\Omega_{r}(r_0)=11.7848$. The oscillating radial motion is slower in the geometry of the solution~\eqref{e14} than it is in the Schwarzschild black hole. In each plot, a perturbed orbit (black plot) oscillates around a stable circular geodesic (magenta or blue plot). The proper time laps between two successive encounters of the black plot and the magenta or blue plot corresponds to a half period of the oscillatory motion. The first encounter takes place in the second quadrant. For the Schwarzschild black hole the second encounter takes place in the fourth quadrant (almost near the $x$-axis) and for the solution~\eqref{e14} the second encounter takes place in the first quadrant.
	
	In Fig.~\ref{figqpos3d} we present a 3D plot depicting the quasi-circular orbit (black plot) obtained upon perturbing a stable circular geodesic (blue plot) in the geometry of the first solution~\eqref{e14}. The perturbed toroidal orbit exhibits both radial and vertical oscillatory motions around the stable blue circle with different epicyclic frequencies $\Omega_{r}$ and $\Omega_\theta$.
	
	In Fig.~\ref{figqpos3dt} we rather choose a circle with radius $r_0$ such that $r_\text{ph}<r_0<r_\text{isco}$. As we know, only a perturbed vertical motion is stable in this region since $\Omega_{\theta}^2>0$ and $\Omega_{r}^2<0$. In the 3D diagram we depict the vertical oscillatory motion (black plot) around a circular orbit (blue plot). It is obvious from the graph that the vertical epicyclic frequency is equal to the orbital frequency: $\Omega_\theta=u^\varphi$. Since there are oscillations in this region, enclosed by $r_\text{ph}$ and $r_\text{isco}$, the energy radiated there should not be neglected.
	
	In this section of numerical analysis we have restricted ourselves to the case of uncharged sources ($Q=0$), the effect of the charge of the source on the perturbed motion is part of the investigations already carried on the Reissner--Nordstr\"om-like and the Kerr--Newman black holes~\cite{qpos3,qposknb}.

	\subsection{Geodesic synchrotron radiations\label{subsecrad}}
	In the previous section while we have considered uncharged sources ($Q=0$), this is however no restriction for the periodic motion of charged test particles is always accompanied by the emission of electromagnetic radiations that can be detected by distant observers. In this section we will not make any restriction and we assume $Q\neq 
	0$.
	
	The total electromagnetic radiation intensity $P_\text{tot}$ radiated by a charged particle of charge $q$ is evaluated following the derivations performed in~\cite{Kerr1,Kerr2,Kerr3}
	\begin{align}\label{P1}
	&P_\text{tot}=\frac{q^2\gamma^4}{6}~\bigg(\frac{3e(r_0)-1}{r_0e(r_0)}\bigg)^2,\\
	&\gamma\equiv \frac{1-2r_0^2\omega^2}{\sqrt{1-3r_0^2\omega^2}}=\frac{1-r_0e'(r_0)}{\big[1-(3/2)r_0e'(r_0)\big]^{1/2}},\nonumber
	\end{align}
	where we have used $\omega^2=(u^{\varphi}/u^t)^2=e'(r_0)/(2r_0)$. This radiation intensity is a factor $1/\gamma^2$ the usual synchrotron radiation intensity due to the ultrarelativistic motion of a charged particle in a magnetic field.
	
	For the first solution~\eqref{e14} and for $c_{13}<0$, the behavior of $P_\text{tot}$ versus the radius of the circle, Fig.~\ref{figrad} (b), almost mimics the behavior of $P_\text{tot}$ for the Schwarzschild metric, Fig.~\ref{figrad} (a): Roughly speaking, the plot first increases and then decreases. The situation is quite different if $c_{13}>0$ where the plot monotonically increases and $P_\text{tot}$ approaches infinity at the point where $\gamma$ diverges. Such an odd behavior where the radiated energy increases monotonically with $r_0$ is rejected physically. Thus, the case with $c_{13}>0$ should be ruled out.
	
	For the second solution~\eqref{e2nd} and for $c_{14}=0$ we observed the same three behaviors as those shown in Fig.~\ref{figrad}. This implies that the case where $c_{14}=0$ and $c_{13}>0$ should be ruled out physically too. The case $c_{14}>0$ is favorable physically for the behavior of $P_\text{tot}$ is similar to that of Fig.~\ref{figrad} (b) and it is independent of the sign of $c_{13}$. In contrast, the case $c_{14}<0$ is not physically favorable for the behavior of $P_\text{tot}$ is similar to that of Fig.~\ref{figrad} (c) regardless of the sign of $c_{13}$ and this case should be ruled out.

	\subsection{Fitting observed resonances\label{subsecres}}
	In the power spectra of Fig.~3 of Ref.~\cite{res} we clearly see two peaks at 300 Hz and 450 Hz representing possible occurrence of the lower $\nu_L=300$ Hz quasi-periodic oscillation (QPO) and of the upper $\nu_U=450$ Hz QPO from the Galactic microquasar GRO J1655-40. Similar peaks have been obtained for the microquasars XTE J1550-564 and GRS 1915+105 obeying the remarkable ratio $\nu_U/\nu_L=3/2$~\cite{qpos1}. Some of the physical properties of these three microquasars and their uncertainties are as follows~\cite{res,res2}:
	\begin{multline}\label{pr1}
	\text{GRO J1655-40 : }\frac{M}{M_\odot}=6.30\pm 0.27,\\
	\nu_U=450\pm 3 \text{ Hz},\;\nu_L=300\pm 5 \text{ Hz},
	\end{multline}
	\begin{multline}\label{pr2}
	\text{XTE J1550-564 : }\frac{M}{M_\odot}=9.1\pm 0.6,\\
	\nu_U=276\pm 3 \text{ Hz},\;\nu_L=184\pm 5 \text{ Hz},
	\end{multline}
	\begin{multline}\label{pr3}
	\text{GRS 1915+105 : }\frac{M}{M_\odot}=14.0\pm 4.4,\\
	\nu_U=168\pm 3 \text{ Hz},\;\nu_L=113\pm 5 \text{ Hz}.
	\end{multline}
	
	These two adjacent or twin values of the QPOs are most certainly due to the phenomenon of resonance which is due to the coupling of non-linear vertical and radial oscillatory motions~\cite{res3,res4}. There are three put-forward models for resonances~\cite{res5}: Parametric resonance, forced resonance and Keplerian resonance. In all three cases, $\nu_U$ and $\nu_L$ are linear combinations of the frequencies $\nu_r$ and $\nu_\theta$ detected by an observer at spatial infinity. These are defined by
	\begin{align}\label{pr4}
	&\nu_r=\frac{1}{2\pi}~\frac{d\tau}{dt}~\Omega_{r}=\frac{1}{2\pi}~\frac{c^2e}{E}~\Omega_{r},\nonumber\\
	&\nu_\theta=\frac{1}{2\pi}~\frac{d\tau}{dt}~\Omega_{\theta}=\frac{1}{2\pi}~\frac{c^2e}{E}~\Omega_{\theta},
	\end{align}
	where we have used $g_{tt} u^{t}  = - E$ [see Eq.~\eqref{phidot}] . Taking $Q=0$ we obtain
	\begin{align}\label{pr5}
	&\nu _r=\frac{c^3 }{2 \pi  G M y^{3/2}} \sqrt{1-\frac{3}{y}-\frac{81 c_{13}}{16 (1-c_{13}) y^4}}\ \times\nonumber\\
	& \sqrt{1+\frac{27 c_{13}}{4 (1-c_{13}) y^3}-\frac{6 [9 c_{13} (3-4 y)-8 (1-c_{13})
			y^3]}{81 c_{13}+16 (1-c_{13}) (3-y) y^3}}\ ,\nonumber\\
	&\nu _{\theta }=\frac{c^3 }{2 \pi  G M y^{3/2}} \sqrt{1+\frac{27 c_{13}}{8 (1-c_{13}) y^3}}\ ,
	\end{align}
	measured in Hz. Here $y\equiv r/r_g$ and $r_g\equiv GM/c^2$. Setting $c_{13}=0$ these expressions reduce to the Schwarzschild ones. It is obvious from the asymptotic behavior at spatial infinity $y\to\infty$ that $\nu _{\theta }>\nu _{r}$. This order remains true near the isco where $\nu _{r}=0$:
	\begin{equation}\label{pr6}
	\nu _{\theta }>\nu _{r}\quad \text{for}\quad y\to\infty\quad \text{and}\quad y\to y_{\text{isco}}\,.
	\end{equation}
	
	Confronting the observed ratio $\nu_U/\nu_L=3/2$ with theory can be done making different assumptions within a resonance model. Said otherwise, in each resonance model~\cite{res5} there may be a set of possible inputs for the same output. Most workers in this field appeal to parametric resonance to explain the observed ratio assuming that $\nu_{\theta }/\nu_{r}=n/2$ and $n\in \mathbb{N}^+$. In almost all applications of parametric resonance one considers the case $n=1$~\cite{b1,b2,b3,b4} where in this case $\nu _{r}$ is the natural frequency of the system and $\nu_{\theta }$ is the parametric excitation ($T_{\theta }=2T_{r}$, the corresponding periods), that is, the vertical oscillations supply energy to the radial oscillations causing resonance~\cite{b4}. However, according to~\eqref{pr6} it is neither possible to have $n=1$ nor $n=2$ in the vicinity of the isco where it is thought that the resonance effects take place. The next allowed choice is thus $n=3$ by which $\nu_{r}$ becomes the parametric excitation that supplies energy to the vertical oscillations. Neglecting the effects of rotation and assuming that $\nu_U=\nu_\theta$, $\nu_L=\nu_r$ along with $\nu_{\theta }/\nu_{r}=3/2$ ($n=3$), it was possible to have good, but partial, curve fits~\cite{fit} to the data of the three microquasars~\eqref{pr1}, \eqref{pr2} and~\eqref{pr3} provided the effects of an external magnetic field on the circular motion of a charged particle are included. In the rightmost panel of Fig.~13 of Ref.~\cite{fit} each curve $\nu_U$ versus $M/M_\odot$ crosses the mass error band of each microquasar for the same value of $\mathcal{B}=0.1$, but nothing is said whether the curve $\nu_L$ versus $M/M_\odot$ crosses the mass error bands drawn at the lower limit values of the three $\nu_L$'s. To the best of our knowledge, there are no curve fits to the data of the three microquasars~\eqref{pr1}, \eqref{pr2} and~\eqref{pr3} if rotation and magnetic fields are ignored. Said otherwise, if the microquasars are described by the Schwarzschild metric, the choice $n=3$ would not lead to any curve fit to the data of the microquasars.
	
	The aim of this section is to show that it is possible to have good and complete curve fits to the data of the three microquasars~\eqref{pr1}, \eqref{pr2} and~\eqref{pr3} even if their rotation is ignored and the influence of any external magnetic field is neglected. We make the following ansatz
	\begin{equation}\label{an}
	\nu_U = 2.6(\nu_\theta-\nu_r),\qquad \nu_L =2.6\nu_r \ .
	\end{equation}
	In parametric resonance model this should correspond to $n=5$; however, there is no need to specify the mechanism or model behind resonance. A feature of the ansatz~\eqref{an} is that the obtained value of $r_0$, solution to $\nu_U/\nu_L=3/2$, is much closer to $r_{\text{isco}}$ where accretion and QPOs occur. 
	
	In Fig.~\ref{figres} the black and blue plots represent $\nu_U$ and $\nu_L$ (in Hz) versus $M/M_\odot$, respectively, and the green lines represent the mass limits as given in~\eqref{pr1}, \eqref{pr2} and~\eqref{pr3}. Each curve (black or blue) crosses the (upper or lower) mass error band of each microquasar for the same value of $c_{13}=-0.6$.
	
	Even if the microquasars are just treated as Schwarzschild black holes, the ansatz~\eqref{an} allows one to provide almost good and complete curve fits to the data of the three microquasars~\eqref{pr1}, \eqref{pr2} and~\eqref{pr3}, as depicted in Fig.~\ref{figresS}.

A word on the ansatz~\eqref{an} is in order. First of all, to the best of our knowledge, in all previous works one relied on the assumptions $\nu_U=\nu_\theta$ and $\nu_L=\nu_r$ for \emph{the sake of simplicity} and one took into account the rotation or magnetic field effects to justify the observed ratio $\nu_U/\nu_L=3/2$. There is no theoretic physical argument to support such assumptions, which were a mere working ansatz, nor to support the ansatz~\eqref{an}. Said otherwise, the ansatz~\eqref{an} is the result of mere empirical observations.

Another empirical argument in favor of the ansatz~\eqref{an} is the extended range of $c_{13}$ yielding good and complete curve fits to the data of the three microquasars~\eqref{pr1}, \eqref{pr2} and~\eqref{pr3}. In Figs.~\ref{figres} and~\ref{figresS} we took $c_{13}=-0.6$ and $c_{13}=0$, respectively. In Fig.~\ref{figres5000} we show that the range of $c_{13}$, in fact, extends down to $-\infty$. In Fig.~\ref{figres5000} we have selected the microquasar GRO J1655-40, which has the narrowest mass band error $\Delta M=0.54 M_\odot$, and we have taken $c_{13}=-5000$. We have observed that as $|c_{13}|$ increases the curve fitting improves. For the other two microquasars, with larger mass band errors, the curve fitting is much better.

	\section{Geodesic equations\label{secgeod}}

	
	We now turn to a non-perturbative study of the geodesic equations. Henceforth, it is convenient to use units where the speed of light equals unity, $c=1$.
	As $\partial_t$ and $\partial_\varphi$ are Killing vectors of the spacetime, we have the first integrals of motion 
	\begin{align}
	u^t=\dot{t}=\frac{E}{e(r)},\quad u^\varphi=\dot{\varphi}=\frac{L}{r^2\sin^2\theta}, \label{phidot}
	\end{align}
	where $E$ and $L$ are conserved quantities used in Sec.~\ref{sececos}, which we interpret as the energy and angular momentum of the particle, respectively.
	
	
	For timelike geodesics, we have $ g_{\mu \nu} \dot x^\mu \dot x^\nu = -1$. Further using Eq.~\Eqref{phidot} to express $\dot t$ and $\dot \varphi$ in terms of $E$ and $L$ we can get
	\bqn
	g_{rr} (u^{r})^2 + g_{\theta \theta} (u^{\theta})^2 &=& -1 - g_{tt} (u^{t})^2  - g_{\varphi\varphi} (u^{\varphi })^2\nb\\
	\frac{\dot{r}^2}{e(r)}+r^2\dot{\theta}^2&=& -1 +\frac{E^2}{e(r)}- \frac{L^2}{r^2\sin^2\theta}.
	\eqn
	
	Since the two black hole solutions are spherically symmetric, we can consider $\theta=\frac{\pi}{2}=\mathrm{constant}$ without loss of generality. Then the above expression can be rearranged into the form
	\bqn
	\dot r ^2 = E^2 - V_{\rm eff}(r), \label{rdot}
	\eqn
	where $\dot r\equiv u^{r}$ and $V_{\rm eff}(r)$ denotes the effective potential and is given by
	\bqn \lb{Veff}
	V_{\rm eff}(r)= \left(1+\frac{L^2}{r^2}\right)e(r).
	\eqn
	One immediately observes that $V_{\rm eff}(r) \to 1$  as $r \to +\infty$, as expected for an asymptotically flat spacetime. With this case, the particles with energy $E >1$ are able to escape to infinity, and $E = 1$ is the critical case between bound and unbound orbits. In this sense, the maximum energy for the bound orbits is $E=1$.
	
	We can obtain the trajectory of a particle by integrating Eqs.~\Eqref{phidot} and \Eqref{rdot} to get $t$, $\varphi$, and $r$ as a function of $\tau$. However, Eq.~\Eqref{rdot} involves taking a square root and it requires that a choice of sign has to be imposed by hand if one were to integrate this equation numerically. Another convenient equation of motion for numerical analysis can be obtained by turning to the $r$-component of the geodesic equation, which is (for $\theta=\frac{\pi}{2}=\mathrm{constant}$)
	\begin{align}
	\ddot{r}&=\frac{e'(r)}{2e(r)}\dot{r}^2-\frac{e'(r)E^2}{2e(r)}+\frac{e(r)L^2}{r^3},
	\end{align}
	where primes denote derivatives with respect to $r$.

\begin{figure}
		\begin{center}
			\includegraphics[width=0.45\textwidth]{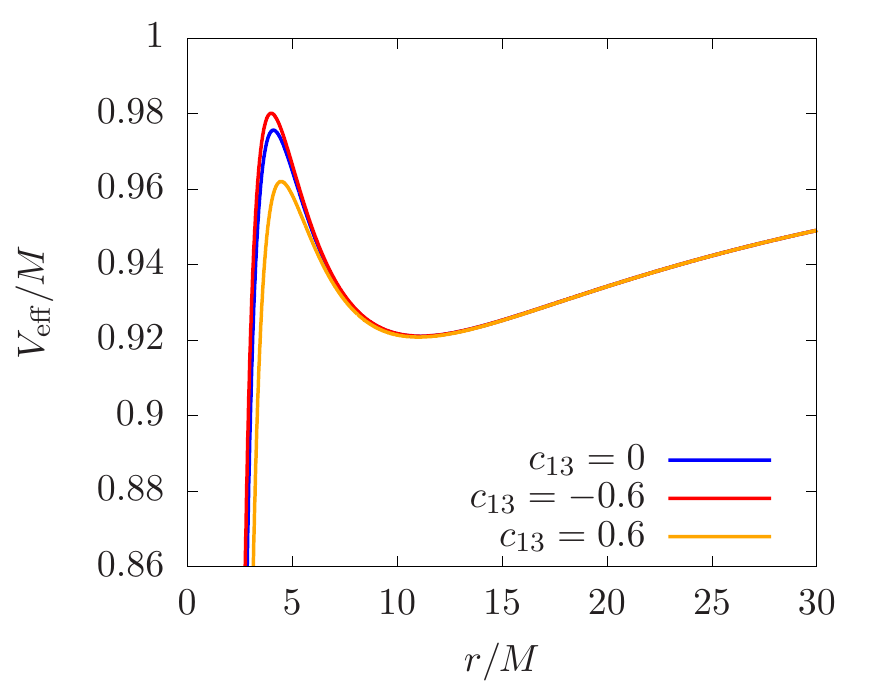}
			\caption{Effective potential for geodesics corresponding to the first solution~\eqref{e14}, for $Q=0$, $L=3.9M$, and various values of $c_{13}$. 
			}
			\label{fig_1stkindVeff}
		\end{center}
		
	\end{figure}

	For the first solution, the plots of the effective potentials for various $c_{13}$ for the uncharged solution are shown in Figs.~\ref{fig_1stkindVeff}. The case $c_{13}=0$ corresponds to the Schwarzchild solution. We see that for the the effect of positive $c_{13}$ is to raise the potential barrier, while negative $c_{13}$ lowers the potential barrier. The effective potential for the second solution shows similar qualitative features, where the details depend on the value of $\frac{3c_{13}-c_{14}}{1-c_{13}}$. We note in passing that typically, for $c_{13}>0$ in the first solution and $\frac{3c_{13}-c_{14}}{1-c_{13}}>0$, there is another turning point of $V_{\mathrm{eff}}$ for small $r$, and hence a potential barrier forms inside the horizon for these two cases. These typically occur at very small $r$ and is not clearly visible in the plots of Figs.~\ref{fig_1stkindVeff}. Similar behavior occurs for effective potentials of the second solution.
	
	For a closer look at the range of allowed orbits, it is convenient to recast the geodesic equations to a new variable. First, we combine \eqref{phidot} and \eqref{rdot}, then we change the radial variable to $r=1/x$. This gives
	\begin{align}
	\frac{d\varphi}{d x}=\pm\frac{L}{\sqrt{P(x)}}, \label{dphidu}
	\end{align}
	where $P(x)$ is a polynomial depending on $e(r)$. Specifically, for the first solution, $P(x)$ is a sixth-degree polynomial, and for the second solution, it is a fourth-degree polynomial. The range of allowed $r$, determined by the condition $\dot{r}^2\geq0$ in Eq.~\Eqref{rdot}, is now translated to ranges of $x$ where $P(x)\geq0$ in Eq.~\Eqref{dphidu}.
	
	In the first solution, where $e(r)$ is given by Eq.~\Eqref{e1st}, $P(x)$ is a sixth-degree polynomial
	\begin{align}
	P(x)&=E^2\nonumber\\
	&\;-\brac{L^2x^2+1}\brac{1-2Mx+Qx^2-16M^4\lambda_0 x^4},
	\end{align}
	where 
	\begin{align}
	\lambda_0&=\frac{27c_{13}}{256(1-c_{13})}. \label{lambda_def}
	\end{align}
	
	By numerical exploration, we find that, for typical parameter ranges relevant to examples in this paper, the equation $P(x)=0$ generically has four real roots and two complex roots. We denote these real roots by $x_\pm$, $x_1$, and $x_2$. For $c_{13}>0$, the roots are ordered as
	\begin{align}
	x_2\leq 0 \leq x_+\leq x_-\leq x_1. \label{x2_left}
	\end{align}
	As $c_{13}\rightarrow0$, the root $x_2$ tends to $-\infty$. This is continued into $c_{13}<0$, where $x_2$ now becomes positive, and the roots are ordered as
	\begin{align}
	x_+\leq x_-\leq x_1\leq x_2. \label{x2_right}
	\end{align}
	For both $c_{13}>0$ and $c_{13}<0$, the range of allowed orbits corresponding to timelike particles in the static Lorentzian region outside the horizon of the black hole are in $x_+\leq x\leq x_-$. Going back to the $r$-coordinates, this corresponds to
	\begin{align}
	r_-=\frac{1}{x_-}\leq r\leq r_+=\frac{1}{x_+}.\label{r_range}
	\end{align}
	
	For the second solution, where $e(r)$ is given by Eq.~\Eqref{e2nd}, $P(x)$ is given by
	\begin{align}
	P(x)&=E^2\nonumber\\
	&\;-\brac{L^2x^2+1}\brac{1-2Mx+\frac{Qx^2}{1-c_{13}}-4M^2\sigma x^2},
	\end{align}
	where
	\begin{align}
	\sigma&=\frac{2c_{13}-c_{14}}{8(1-c_{13})}. \label{sigma_def}
	\end{align}
	
	In this case, $P(x)$ is a fourth-order polynomial, and its roots of $P(x)=0$ can be obtained exactly. Furthermore, the fact that $P(x)$ is a fourth-degree polynomial allows the geodesic equations to be solved exactly. Following the convention of the first solution, we denote the roots by $x_\pm$, $x_1$, and $x_2$. For $\sigma=\frac{2c_{13}-c_{14}}{8(1-c_{13})}>0$, the roots are ordered as Eq.~\Eqref{x2_left}, and the geodesic equation can be solved to obtain $\varphi$ as a function of inverse radius as
	\begin{align}
	\varphi(x)&=\varphi_0\nonumber\\
	&\hspace{-0.5cm}+\int_{x_+}^x\frac{dx'}{2M\sqrt{\sigma (x_1-x')(x_--x')(x'-x_+)(x'-x_2)}}.
	\end{align}
	This integral can be evaluated exactly to give (see 3.147--5 of \cite{gradshteyn2014table})
	\begin{align}
	\varphi(x)&=\varphi_0+\frac{\mathrm{F}\brac{\xi_1,\eta_1}}{M\sqrt{\sigma(x_--x_+)(x_--x_2)}}, \label{2ndkind_exacta}
	\end{align}
	where $\mathrm{F}(\xi,\eta)$ is the elliptic function of the first kind, and
	\begin{align}
	\xi_1&=\arcsin\sqrt{\frac{(x_1-x_+)(x_--x)}{(x_--x_+)(x_1-x)}},\nonumber\\
	\eta_1&=\sqrt{\frac{(x_--x_+)(x_--x_2)}{(x_1-x_+)(x_--x_2)}}.
	\end{align}

	On the other hand, for $\sigma=\frac{2c_{13}-c_{14}}{8(1-c_{13})}<0$, the roots are ordered as \Eqref{x2_right}. The solution of the geodesic equations are then
	\begin{align}
	\varphi(x)&=\phi_0\nonumber\\
	&\hspace{-0.6cm}+\int_{x_+}^x\frac{dx'}{2M\sqrt{|\sigma| (x_2-x')(x_1-x')(x_--x')(x'-x_+)}}.
	\end{align}
	This integral is evaluated exactly to give (see 3.147--3 of \cite{gradshteyn2014table})
	\begin{align}
	\varphi(u)&=\phi_0+\frac{\mathrm{F}\brac{\xi_2,\eta_2}}{M\sqrt{|\sigma|(x_2-x_-)(x_1-x_+)}}, \label{2ndkind_exactb}
	\end{align}
	where
	\begin{align}
	\xi_2&=\arcsin\sqrt{\frac{(x_2-x_1)(x-x_+)}{(x_--x_+)(x_2-x)}},\nonumber\\
	\eta_2&=\sqrt{\frac{(x_2-x_1)(x_--x_+)}{(x_2-x_-)(x_1-x_+)}}.
	\end{align}

	\section{Marginally bound orbits and ISCOs\label{secmb}}
	
	\subsection{Marginally bound orbits}
	
	The equatorial circular orbits are corresponding to those orbits with constant $r$, i.e., $\dot r^2=0$. For these orbits the marginally bound orbits are defined by the following conditions,
	\bqn\lb{mar_condition}
	\dot r^2=E- V_{\rm eff}=0,\;\; \frac{dV_{\rm eff}(r)}{dr}=0,
	\eqn
	with $E=1$. Using the first equation in~\eqref{EL} and solving $E=1$ for $e'(r_{\rm mb})$ we obtain
	\bqn\lb{mar_radius}
	e'( r_{\rm mb}) = \frac{2[1- e(r_{\rm mb})]e(r_{\rm mb})}{ r_{\rm mb}},
	\eqn
	which is the equation satisfied by $r_{\rm mb}$. Using this in the second equation in~\eqref{EL} we obtain
	\bqn\lb{lmb}
	L_{\rm mb} =  \pm r_{\rm mb}\sqrt{\frac{ [1-e( r_{\rm mb})]}{e( r_{\rm mb})}}\,.
	\eqn
	
\begin{figure*}[!htb]
		\includegraphics[width=0.49\textwidth]{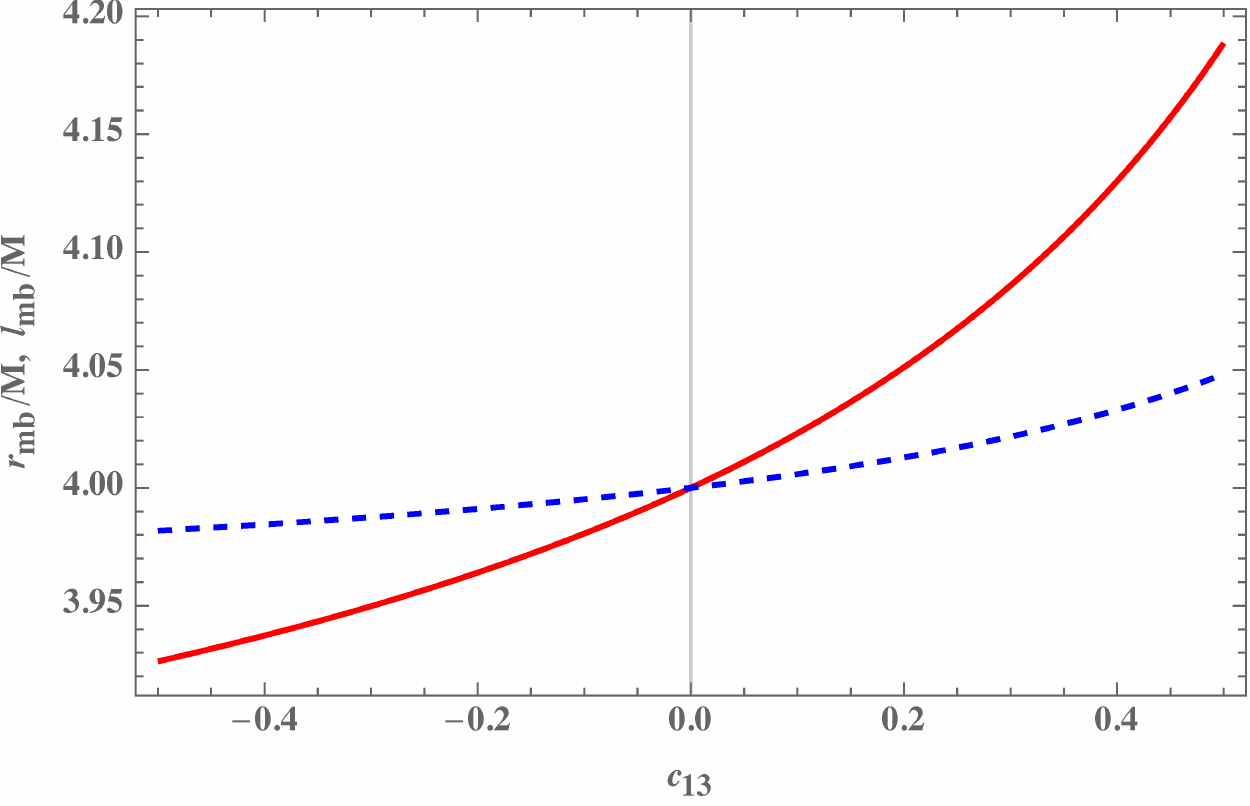}
		\includegraphics[width=0.49\textwidth]{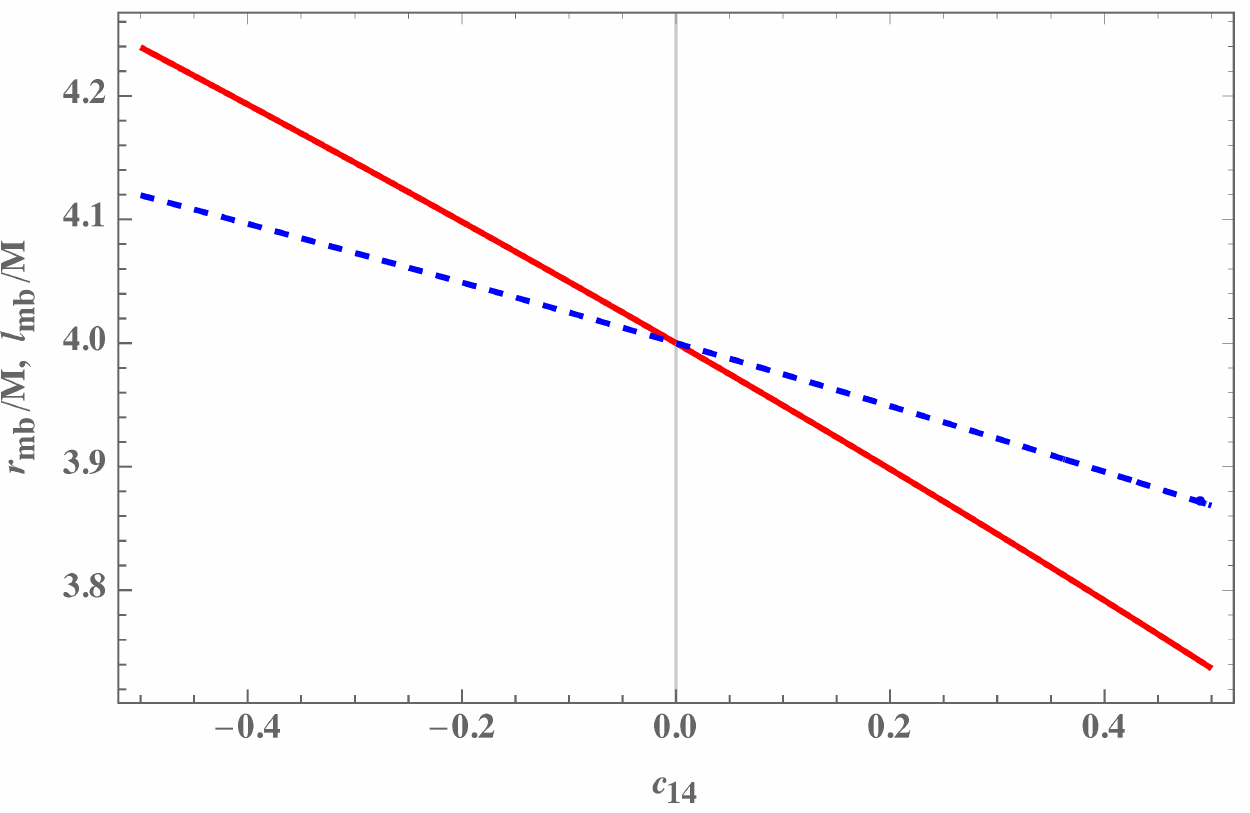}
		\caption{The radius (red solid line) and the angular momentum (blue dashed line) at the marginally bound orbits for the two types of the Einstein-\AE{}ther black holes. Left panel: the first type neutral black hole ($c_{14}=0$ but $c_{123}\neq 0$, Q=0). Right panel: the second type neutral black hole ($c_{123} =0, c_{13}=0, Q=0$).} \label{mb_rl1}
	\end{figure*}

	
	For the first type Einstein-\AE{}ther black hole ($c_{14}=0$ but $c_{123} \neq 0$), the condition (\ref{mar_radius}) with $e(r)$ given by (\ref{e14}) reduces to
	\bqn
	&&\frac{2(Q-2 Mr-2 r^2)}{r^3} - \frac{32 \lambda_0 M^4}{r^5} \nb\\
	&&+ \frac{2 r (Q-3 M r +2 r^2)}{r^2 (Q- 2 M r + 2 r^2)- 16 \lambda_0 M^4} = 0,
	\eqn
	where $\lambda_0$ is as given in Eq.~\Eqref{lambda_def}.
	This equation can only be solved numerically. With a given value of $c_{13}$, one can solve the above equation and obtain the value of $r_{\rm mb}$ and $L_{\rm mb}$, which are presented in the left panel of Fig.~\ref{mb_rl1} for a neutral black hole ($Q=0$).
	
	\begin{figure*}[!htb]
		\includegraphics[width=0.49\textwidth]{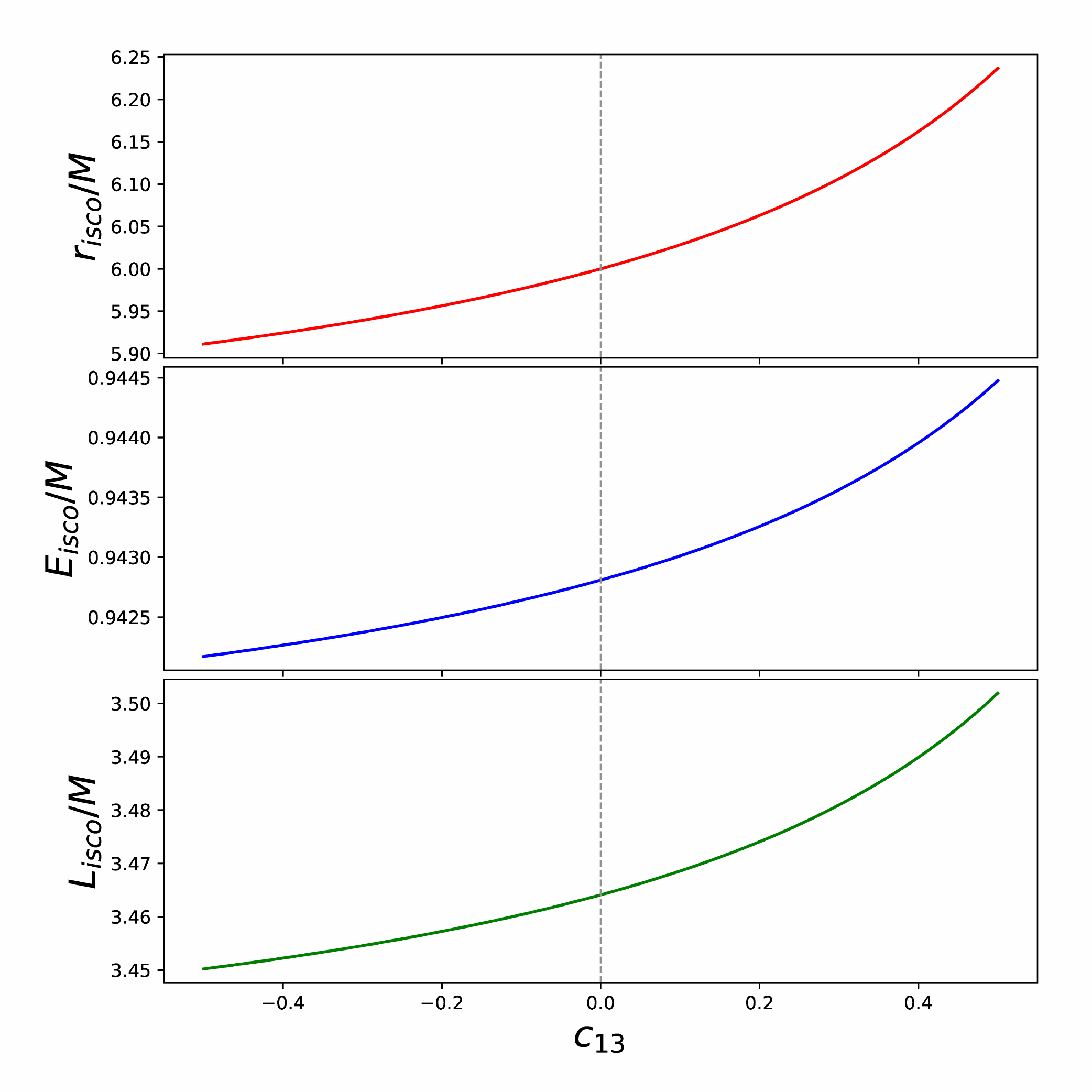}
		\includegraphics[width=0.49\textwidth]{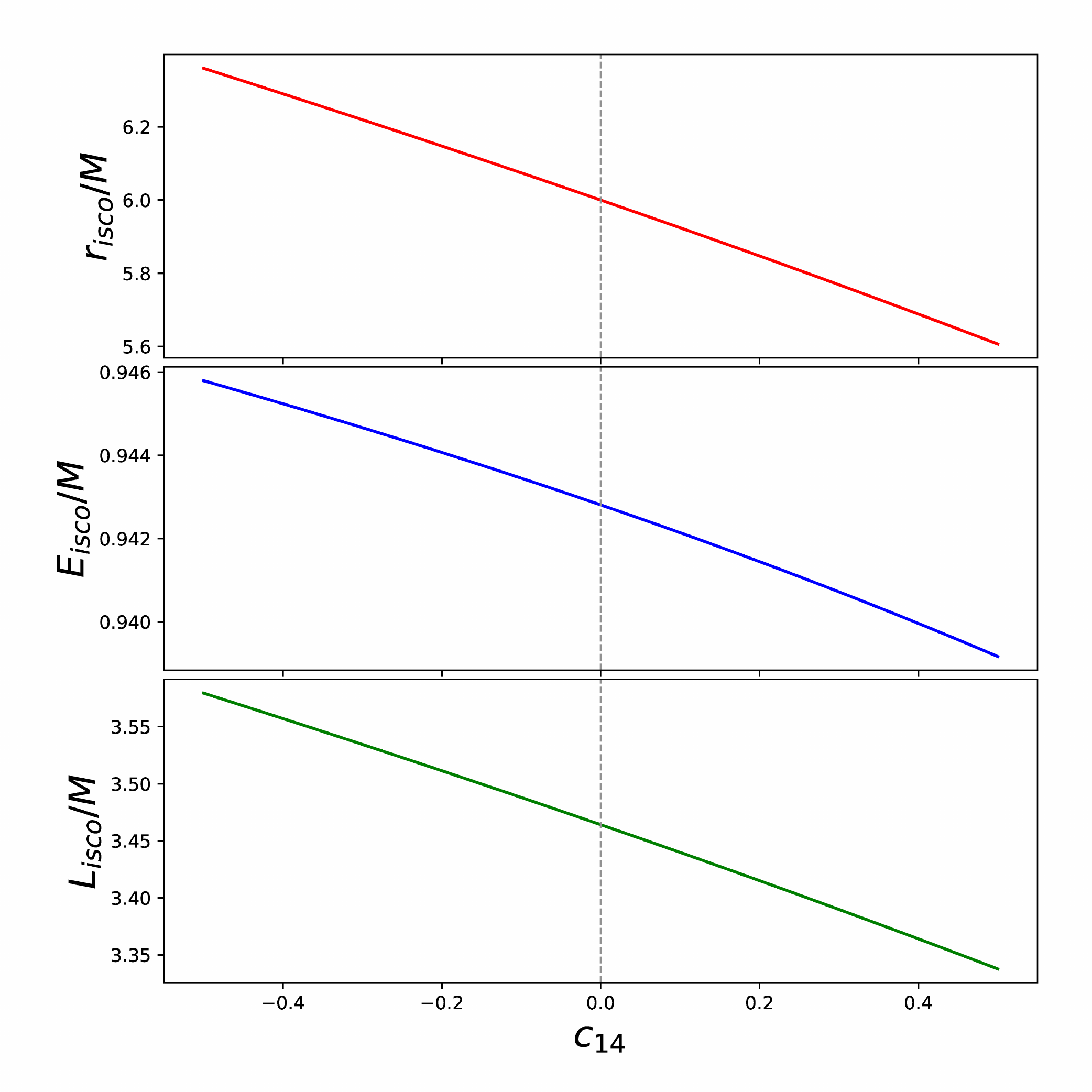}
		\caption{The radius $r_{\rm isco}$, energy $E_{\rm isco}$, and the angular momentum $L_{\rm isco}$ at the innermost stable orbits for the two types of the Einstein-\AE{}ther black holes respectively. Left panel: the first type neutral black hole ($c_{14}=0$ but $c_{123}\neq 0$, Q=0). Right panel: the second type neutral black hole ($c_{123} =0, c_{13}=0, Q=0$).} \label{ISCO}
	\end{figure*}
	\begin{figure*}[!htb]
		\includegraphics[width=0.49\textwidth]{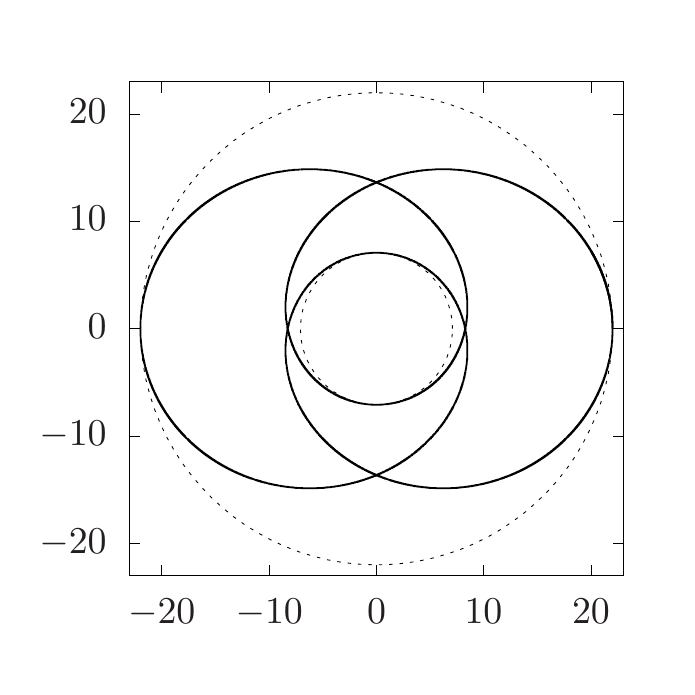}
		\includegraphics[width=0.49\textwidth]{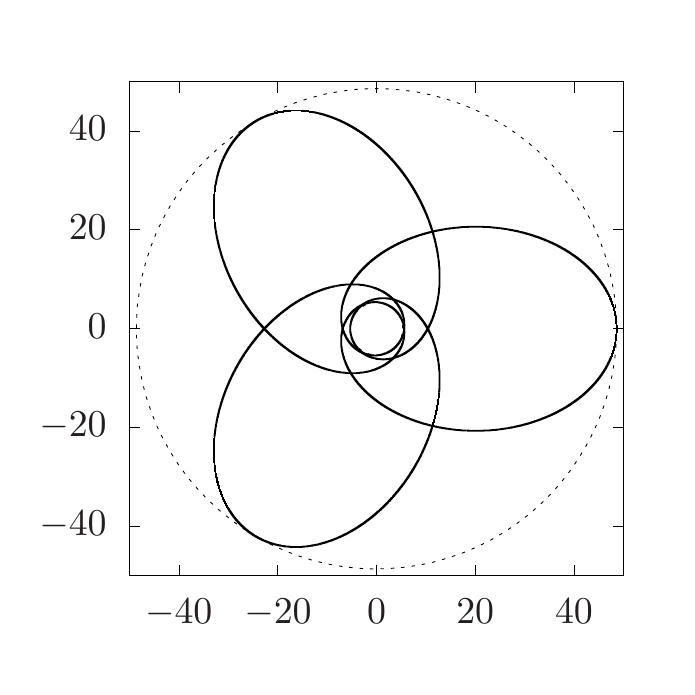}
		\includegraphics[width=0.49\textwidth]{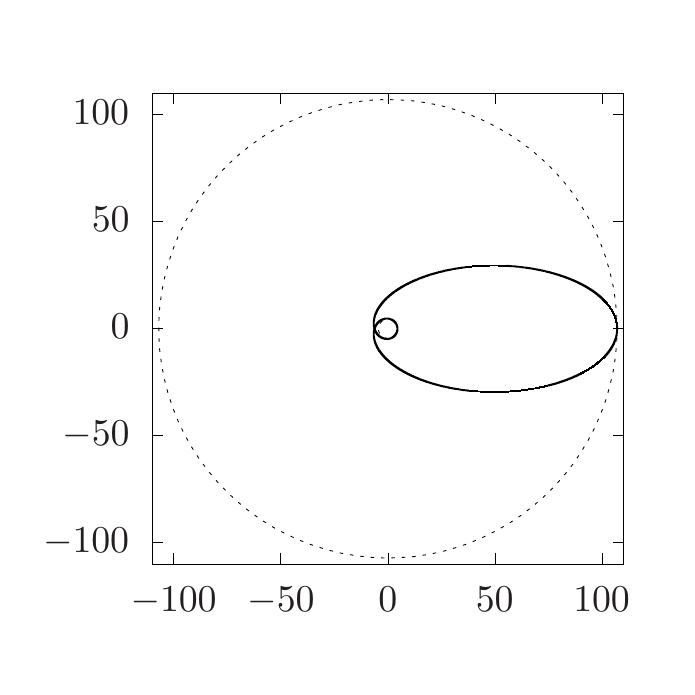}
		\includegraphics[width=0.49\textwidth]{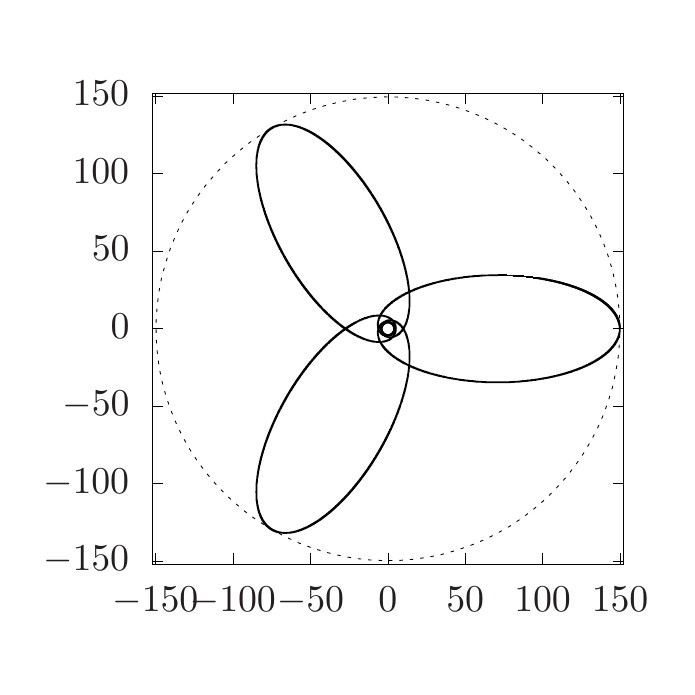}
		\caption{Periodic orbits, plotted in Cartesian-like coordinates $X=r\cos\varphi$ (horizontal axis) and $Y=r\sin\varphi$ (vertical axis) in units of $M$ for $Q=0$, $c_{14}=0.2$, $L=3.9$. The values of $(z,w,v)$ and $E$ for each orbits are, clockwise from top left, the $(2,0,1)$-orbit with $E=0.968428$, $(3,0,2)$-orbit with $E=0.9823845$, the $(1,1,0)$ orbit with $E=0.99127729$, and the $(3,1,1)$-orbit with $E=0.99363628$.}
		\label{fig_2ndPeriodic}
	\end{figure*}

	For the second type Einstein-{\AE}ther black hole ($c_{123} =0$), the condition (\ref{mar_radius}) with $e(r)$ given by (\ref{e123}) reduces to
	\bqn
	\left(\frac{r}{M}\right)^3- 4 \left(\frac{r}{M}\right)^2 + 4 \frac{Q_{14}}{M^2} \frac{r}{M} - \frac{Q_{14}^2}{M^4}=0,
	\eqn
	where
	\bqn
	Q_{14} \equiv \frac{1}{1-c_{13}} \left(Q- \frac{2c_{13} - c_{14}}{2}\right).
	\eqn
	This equation can be solved analytically yielding
	\bqn
	r_{\rm mb} = \frac{4M}{3}- \frac{\sqrt[3]{2} (12 M^2 Q_{14} - 16 M^4)}{3 M \mathcal{Y}_1} +\frac{\mathcal{Y}_1}{3\sqrt[3]{2} M},\nb\\
	\eqn
	where
	\bqn
	\mathcal{Y}_1&\equiv& \Big(128M^6- 144 M^4 Q_{14} + 27 M^2 Q_{14}^2 \nb\\
	&&~~+ 3 \sqrt{3} \sqrt{27M^4 Q_{14}^4- 32 M^6 Q_{14}^3}\Big)^{1/3}.
	\eqn
	With $r_{\rm mb}$, the angular momentum $L_{\rm mb}$ at the marginally bound orbit can be calculated via (\ref{lmb}). In the right panel of the Fig.~\ref{mb_rl1}, we present the results of $(r_{\rm mb}, L_{\rm mb})$ with respect to $c_{14}$ for a neutral black hole ($Q=0$) and $c_{13}=0$.

	\subsection{Innermost stable circular orbits}
	
	As we mentioned in the above, the marginally bound orbit corresponds to the bound orbit that has the maximum energy $E=1$. All the bound orbits which have energy $E<1$ can only exist beyond $r_{\rm mb}$, i.e, $r>r_{\rm mb}$. The stability of these orbits are determined by the sign of $d^2 V_{\rm eff}(r)/dr^2$. As we have seen in Sec.~\ref{ECOs}, it is enough to impose the condition $\Omega_{r}^2>0$ to ensure the radial and vertical stability of circular orbits and this corresponds to $d^2 V_{\rm eff}(r)/dr^2 >0$~\eqref{f4}. Consequently, unstable circular orbits have $d^2 V_{\rm eff}(r)/dr^2<0$. The critical condition,
	\bqn\lb{third}
	\Omega_{r}(r_{\rm isco})^2=\frac{d^2V_{\rm eff}(r)}{dr^2}\Big|_{r=r_{\rm isco}}=0,
	\eqn
	together with the conditions in (\ref{mar_condition}) for $E<1$ determine the radius of the innermost stable circular orbit. This amounts to solve Eq.~\eqref{r_isco}. The energy and angular momentum ($E_{\rm isco},\,L_{\rm isco}$)	are given by~\eqref{EL} on replacing $r_0$ by $r_{\rm isco}$
	\bqn
	E_{\rm isco} &=& e(r_{\rm isco})\sqrt{\frac{2}{2e(r_{\rm isco})-r_{\rm isco}e'(r_{\rm isco})}}\,, \lb{Eisco}\\
	L_{\rm isco} &=& \pm r_{\rm isco}^{3/2}\sqrt{\frac{e'(r_{\rm isco})}{2e(r_{\rm isco})-r_{\rm isco}e'(r_{\rm isco})}}\,. \lb{lisco}
	\eqn
	
	Using these relations we discuss separately the isco for the two types of Einstein-{\AE}ther black holes.
	
	\subsubsection{ISCO for first type black hole (\texorpdfstring{$c_{14}=0$}{c14=0} but \texorpdfstring{$c_{123} \neq 0$}{c123=/=0})}
	
	For the first type black hole, the function $e(r)$ is given by (\ref{e14}). In this case, Eq.~(\ref{r_isco}) can not be solved analytically. In Fig.~\ref{ISCO}, we plot the results of $r_{\rm isco}$, $E_{\rm isco}$, and $L_{\rm isco}$ with respect to the {\ae}ther parameter $c_{13}$ for the first type neutral black hole ($Q=0$). It is shown that the radius, energy, and angular momentum for the isco all increase with $c_{13}$. When the {\ae}ther field is absent (i.e. $c_{13}=0$), all these quantities reduce to those of the Schwarschild black hole.

	\subsubsection{ISCO for second type black hole (\texorpdfstring{$c_{123}= 0$}{c123=0})}
	
	For the second type black hole, the function $e(r)$ is given by (\ref{e123}). In this case, Eq.~(\ref{r_isco}) reduces to
	\bqn
	\frac{3 (M r_{\rm isco} -Q_{14})(r_{\rm isco}^2 - 2 M r_{\rm isco} +Q_{14})}{2 M r_{\rm isco}^3 - 3 r_{\rm isco}^2 Q_{14} +Q_{14}^2}=1,
	\eqn
	which leads to
	\bqn
	r_{\rm isco} = 2 M +\mathcal{Y}_2+ \frac{4 M^2- 3 Q_{14}}{\mathcal{Y}_2},
	\eqn
	where
	\bqn
	\mathcal{Y}_2 &\equiv& \Bigg(\frac{8M^4 -9M^2 Q_{14} +2 Q_{14}^2}{M} \nb\\
	&&~~+\frac{Q_{14}\sqrt{5 M^4 - 9M^2 Q_{14} + 4 Q_{14}^2}}{M}\Bigg)^{1/3}.
	\eqn
	The energy and the angular momentum can be calculated from (\ref{Eisco}) and (\ref{lisco}). In Fig.~\ref{ISCO}, we plot the results of $r_{\rm isco}$, $E_{\rm isco}$, and $L_{\rm isco}$ with respect to the {\ae}ther parameter $c_{14}$ for the second type neutral black hole ($Q=0$) by setting $c_{13}=0$. It is shown that the radius, energy, and angular momentum for the isco all decrease with $c_{14}$. When the {\ae}ther field is absent (i.e. $c_{14}=0=c_{13}$), all these quantities reduce to those of the Schwarschild black hole.
	
	\section{Periodic orbits\label{secpo}}

	In this section, we shall seek periodic timelike orbits around the Einstein-{\AE}ther black holes. We adopt taxonomy as introduced in \cite{Levin:2008mq} for indexing all periodic orbits around the Einstein-{\AE}ther black holes with a triplet of integers $(z, w, v)$, which  describe the zoom ($z$), whirl ($w$), and vertex ($v$) behaviors. Periodic orbits are defined as orbits that return exactly to their initial conditions after a finite time. Viewing $r$ and $\varphi$ as functions of the affine parameter $\tau$, periodic orbits require that the ratio between the two frequencies of oscillations in the $r(\tau)$ and $\varphi(\tau)$-motion to be a rational number. 
	
	As detailed in \cite{Levin:2008mq}, a generic aperiodic orbit around the black hole can be approximated by a nearby periodic orbit since any irrational number can be approximated by a nearby rational number. Therefore, the exploration of the periodic orbits would be very helpful for understanding the structure of any generic orbits and the corresponding radiation of the gravitational waves.
	
	According to the taxonomy of ref.~\cite{Levin:2008mq}, we introduce the ratio $q$ between the two frequencies, $\omega_r$ and $\omega_\varphi$ of oscillations in the $r(\tau)$ and $\varphi(\tau)$-motion respectively, in terms of three integers $(z, w, v)$ as
	\bqn
	q\equiv \frac{\omega_\varphi}{\omega_r}-1 = w + \frac{v}{z}.
	\eqn
	Here $\frac{\omega_\varphi}{\omega_r}=\Delta \varphi/(2\pi)$ with $\Delta \varphi \equiv \oint d\varphi$ being the equatorial angle during one period in $r$, which is required to be an integer multiple of $2\pi$. Using the geodesic equations of the Einstein-{\AE}ther black hole, $q$ can be calculated via
	\begin{align}
	q=\frac{1}{\pi}\Delta\varphi-1,
	\end{align}
	where 
	\begin{align}
	\Delta\varphi=\int_{x_+}^{x_-}\frac{L\, dx'}{\sqrt{P(x')}}.
	\end{align}
	As we have shown in Sec.~\ref{secgeod}, in the case of the second-type Einstein-{\AE}ther black hole, $P(x)$ is a fourth-order polynomial and can be integrated exactly and is given by Eqs.~(\ref{2ndkind_exacta}) for positive $\sigma$ and Eq.~(\ref{2ndkind_exactb}) for negative $\sigma$. For the first-type Einstein-{\AE}ther black hole, $P(x)$ is a sixth-order polynomial and can be integrated numerically. In the present section, we are looking specifically for solutions which produces rational $q$, corresponding to periodic orbits. Some examples periodic orbits are shown in Fig.~\ref{fig_2ndPeriodic}.
	

	
\begin{figure}
		\begin{center}
			\includegraphics[width=0.48\textwidth]{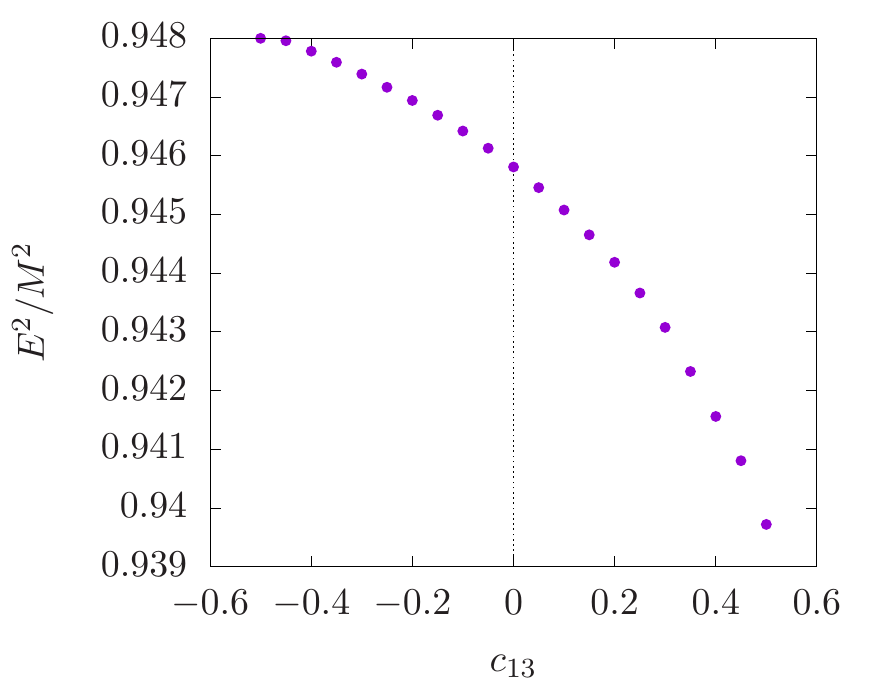}
			\caption{Values of $E^2$ vs $c_{13}$ for the $(5,0,3)$ periodic orbit, in units of $M$ where $q=0.6$, for the second solution with $L=3.9$, $c_{13}=0$, and $Q=0$.}
			\label{fig_changec13}
		\end{center}
	\end{figure}

\begin{figure}
		\begin{center}
			\includegraphics[width=0.48\textwidth]{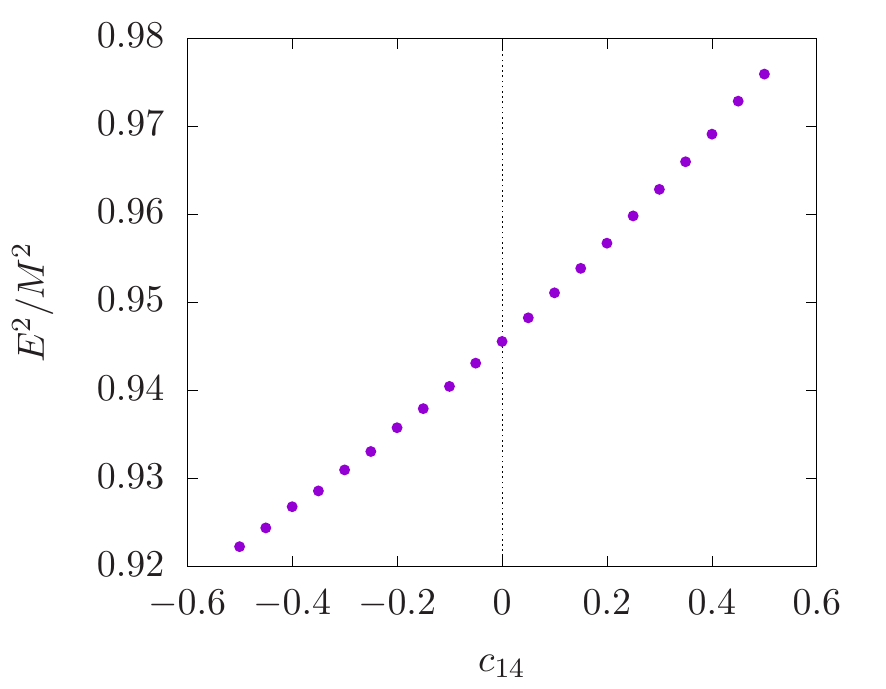}
			\caption{Values of $E^2$ vs $c_{14}$ for the $(5,0,3)$ periodic orbit, in units of $M$ where $q=0.6$, for the second solution with $L=3.9$, $c_{14}=0$, and $Q=0$.}
			\label{fig_changec14}
		\end{center}
	\end{figure}

	We find that the geometry of periodic orbits of both types of Einstein-{\AE}ther black holes are similar to the periodic orbits in the Schwarzschild case, and hence we can adopt the same $(z,w,v)$-taxonomy of \cite{Levin:2008mq}. However, the orbital parameters $(E,L)$ for a particular $(z,w,v)$ orbits differ from the standard Schwarzschild case. For concreteness, let us choose to study the $(5,0,3)$, which corresponds to $q=0.6$. We shall also fix the angular momentum to be $L=3.9M$ (following the examples in Fig.~11 of \cite{Levin:2008mq}). As $c_{13}$ and $c_{14}$ changes, the required energy $E$ deviates from its corresponding Schwarzschild value. For the first type of Einstein-{\AE}ther black hole,  Fig.~\ref{fig_changec13} shows how the energy for the $(5,0,3)$ orbit changes with $\lambda=\frac{27c_{13}}{256(1-c_{13})}$.
	On the other hand, for the second type of Einstein-{\AE}ther black hole, Fig.~\ref{fig_changec14} shows how the energy of the $(5,0,3)$ orbit changes with $\sigma=\frac{2c_{13}-c_{14}}{8(1-c_{13})}$. For the figure, we fixed $c_{13}=0$, and varied $c_{14}$.
	
\begin{figure*}
		\begin{center}
			\includegraphics[width=0.49\textwidth]{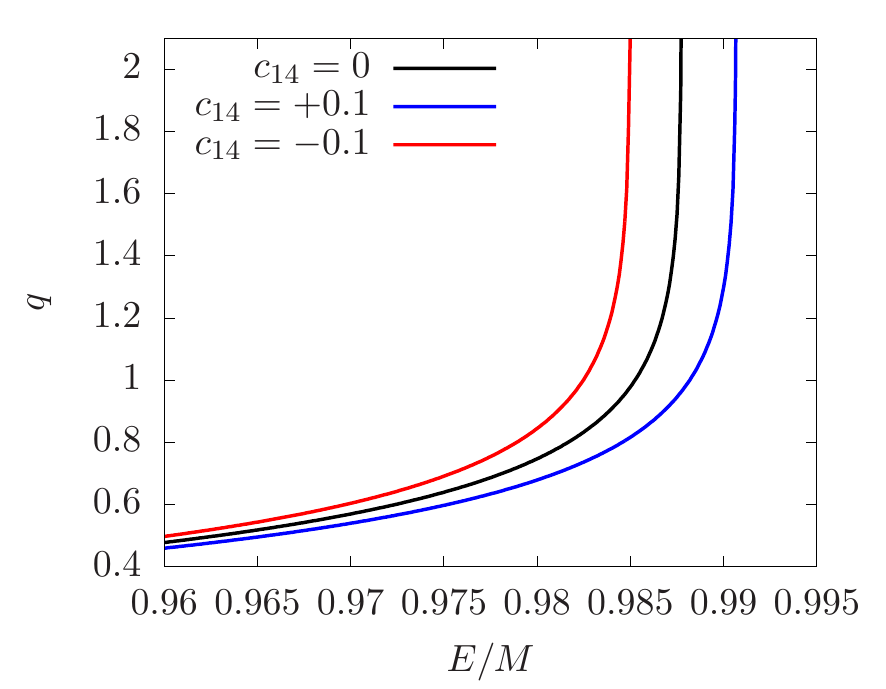}
			\includegraphics[width=0.49\textwidth]{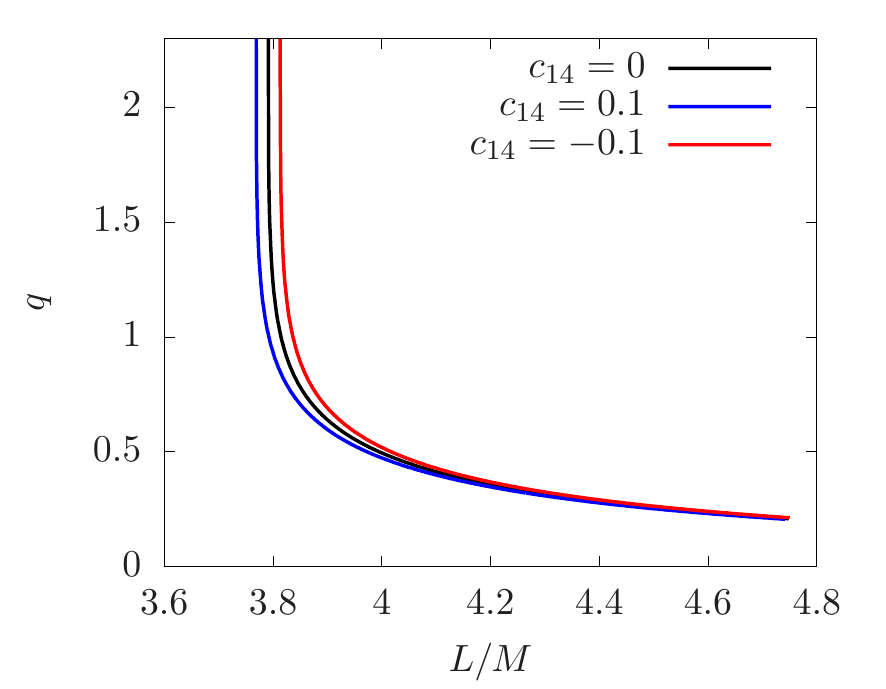}
			\caption{Left: Plots of $q$ vs $E$ for the second solution with $Q=0$ and $L=3.9M$. Right: Plot of $q$ vs $L$ for the second solution with $Q=0$, and $E=0.975M$.}
			\label{fig_qEqL}
		\end{center}
	\end{figure*}

	We can further explore how $q$ varies with $E$ and $L$ for the second solution of the Einstein-{\AE}ther black hole. We can compare orbits with the same energy and angular momentum with the Schwarzschild case ($c_{13}=c_{14}=0$). Orbits with positive $\sigma$ will have a smaller $q$ compared to the Schwarzschild case (the black curve in Fig.~\ref{fig_qEqL}). This means, if we were to observe an orbit of some given $E$ and $L$ of the second-type Einstein-{\AE}ther black hole, it will undershoot the value predicted by standard Einstein gravity. Similarly, for negative $\sigma$, it will overshoot the standard prediction. The same conclusion holds for the first solution of the Einstein-{\AE}ther black hole, with positive $\lambda$ undershooting the standard prediction and a negative $\lambda$ overshooting it.   
	
	Information about periodic orbits may provide insights to the phenomena of black hole mergers with extreme mass ratios. Specifically, the motion of the smaller black hole (or another compact object such as a neutron star) around a larger one may be approximated by a test particle trajectory around a central black hole. The falling of the smaller black hole towards the larger one can be viewed as a sequence of transitions between periodic orbits, in which energy and angular momentum are emitted in gravitational waves \cite{Levin:2008mq,Levin:2009sk}.
	
	As the smaller object transitions across different orbits, we may write its rate of change of $q$ as
	\begin{align}
	\frac{dq}{dt}=\frac{\partial q}{\partial E}\frac{d E}{dt}+\frac{\partial q}{\partial L}\frac{d L}{dt}.
	\end{align}
	Of particular interest is the resonance during inspiral, which corresponds to $\frac{dq}{dt}\simeq 0$. Clearly, such a phenomenon is possible if $\frac{\partial q}{\partial E}$ and $\frac{\partial q}{\partial L}$ have opposite signs. For the case of Einstein-{\AE}ther black holes, we can check from the slopes of $q$ vs $E$ and $q$ vs $L$, such as in Fig.~\ref{fig_qEqL}, we see that in-spiraling resonance is indeed possible.

	\section{Null geodesics\label{nullgeo}}

\begin{figure}
		\begin{center}
			\includegraphics[width=0.49\textwidth]{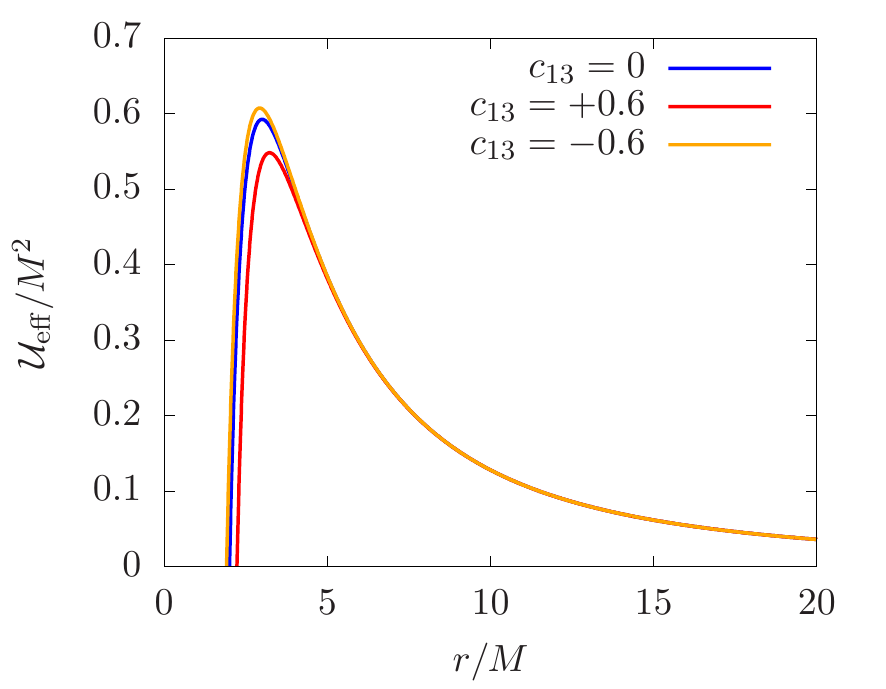}
			\caption{The effective potential for null geodesics in the first black-hole solution~\eqref{e14} in units of $M$ for $Q=0$ and $L=4M$. 
			}
			\label{fig_VeffNull}
		\end{center}
	\end{figure}

	For the case of null geodesics, we have $g_{\mu\nu}\dot{x}^\mu\dot{x}^\nu=0$, and the geodesic equations (for $\theta=\frac{\pi}{2}=\mathrm{constant}$) are 
	\begin{align}
	\dot{t}&=\frac{E}{e(r)},\\
	\dot{\varphi}&=\frac{L}{r^2},\label{ph_phidot}\\
	\dot{r}^2&=E^2-\mathcal{U}_{\mathrm{eff}},\label{ph_rdot},
	\end{align}
	where the effective potential for photon orbits are
	\begin{align}
	\mathcal{U}_{\mathrm{eff}}=\frac{L^2}{r^2}e(r). \label{ph_eff_potential}
	\end{align}
	The effective potential of the first type~\eqref{e14} of black holes for null geodesics is shown in Fig.~\ref{fig_VeffNull}. For the second solution~\eqref{e2nd} we have obtained similar plots. They have the same qualitative shape as for Schwarzschild null geodesics, though the radii of their photon spheres (represented by location of turning points of $\mathcal{U}_{\mathrm{eff}}$) varies with $c_{13}$ and $c_{14}$. 
	
	In the following, we shall use these equations to obtain circular photon orbits and to calculate the bending angles for gravitational lensing.	

	\subsection{Circular photon orbits}
	From Eq.~\eqref{ph_rdot}, we can easily obtain the equations of circular photon orbits by solving $\frac{d \mathcal{U}_{\mathrm{eff}}}{dr}=0$. For both types of Einstein-{\AE}ther black hole, we determine unstable circular photon orbits.
	
	\emph{First solution} --- For the first solution of the Einstein-{\AE}ther black hole, the equation $\frac{d \mathcal{U}_{\mathrm{eff}}}{dr}=0$ leads to 
	\begin{align}
	r^4-3Mr^3-48\lambda M^4=0,
	\end{align}
	where $\lambda$ is as defined in Eq.~\eqref{lambda_def}. By numerical exploration, we find that for typical parameter ranges of the first solution, this equation has two real roots. The larger one, $r_{\mathrm{ph}}$, is located outside the horizon. Furthermore, evaluating the second derivative of the effective potential gives 
	\begin{equation}
	\frac{d^2\mathcal{U}_{\mathrm{eff}}}{dr^2}\bigg|_{r_{\mathrm{ph}}}<0,
	\end{equation}
	showing that the circular photon orbits are unstable. 
	
	\emph{Second solution} --- For the second solution of the Einstein-{\AE}ther black hole, the equation $\frac{d \mathcal{U}_{\mathrm{eff}}}{dr}=0$ is solved by
	\begin{align}
	r_{\mathrm{ph}\pm}=\frac{1}{2}\brac{3\pm\sqrt{9+32\sigma}}M,
	\end{align}
	where $\sigma$ is given in Eq.~\Eqref{sigma_def}. Typically, only $r_{\mathrm{ph}+}$ is located outside the horizon. Recalling that the radius of the photon sphere for the Schwarzschild spacetime is $r=3M$, we find that the photon sphere is larger than the Schwarzchild photon sphere for $\sigma>0$, and smaller for $\sigma<0$.
	
	Evaluating the second derivative of $\mathcal{U}_{\mathrm{eff}}$ at the photon sphere leads to 
	\begin{equation}
	\frac{d^2\mathcal{U}_{\mathrm{eff}}}{dr^2}\bigg|_{r_{\mathrm{ph}+}}=-\frac{64L^2\sqrt{9+32\sigma}}{M^4\brac{3+\sqrt{9+32\sigma}}^5}<0,
	\end{equation}
	which is negative, indicating that the circular orbits are unstable.

	\subsection{Gravitational lensing by the Einstein-{\AE}ther black hole}
	
	From Eq.~\Eqref{ph_phidot} and \Eqref{ph_rdot}, we have
	\begin{align}
	\frac{d\varphi}{dx}&=\frac{L}{\sqrt{E^2-L^2x^2e(1/x)}},\label{ph_dphidu1}
	\end{align}
	where, again, we have introduced the coordinate transformation $x=1/r$.

	As in the case of timelike geodesics, the range of allowed $r=1/x$ for photon geodesics are determined by the requirement that $\dot{r}^2\geq0$ in Eq.~\Eqref{ph_rdot}. For typical parameter ranges of the two types of Einstein-{\AE}ther black holes, a generic photon trajectory which does not fall into the black hole will lie in a range 
	\begin{align}
	0\leq x\leq x_0\quad\leftrightarrow\quad \infty\geq r\geq r_0,
	\end{align}
	where $r_0=\frac{1}{x_0}$ is a root of $\dot{r}=0$, and corresponds to the (coordinate) distance of closest approach to the black hole. Observe that for the case of photon geodesics, the constants of motion always appear together in a ratio $E/L$, and this can be expressed in terms of $r_0=1/x_0$ using Eq.~\Eqref{ph_rdot} as 
	\begin{align}
	\frac{E}{L}=x_0\sqrt{e(1/x_0)}.
	\end{align}
	Therefore, it is convenient to parametrize the trajectories by a single parameter $x_0$. In terms of this parameter, Eq.~\Eqref{ph_dphidu1} is written as 
	\begin{align}
	\frac{d\varphi}{dx}&=\frac{1}{\sqrt{G(x)}}. \label{ph_dphidu2}
	\end{align} 
	where $G(x)$ is given by 
	\begin{align}
	G(x)=x_0^2e(1/x_0)-x^2e(1/x).
	\end{align}
	We are interested in the case of photon trajectories from infinity being deflected by the Einstein-{\AE}ther black hole. Furthermore, let us consider the uncharged case $Q=0$ as this is the case with the most astrophysical relevance. Since the spacetime is asymptotically flat, the change in coordinate angle $\varphi$ gives an accurate depiction of the bending angle of light. By integrating \Eqref{ph_dphidu2},
	\begin{align}
	\psi\equiv2\Delta\varphi=2\int_0^{x_0}\frac{d x'}{\sqrt{G(x)}}. \label{ph_dphidu3} 
	\end{align}
	
\begin{figure}
		\begin{center}
			\includegraphics[width=0.48\textwidth]{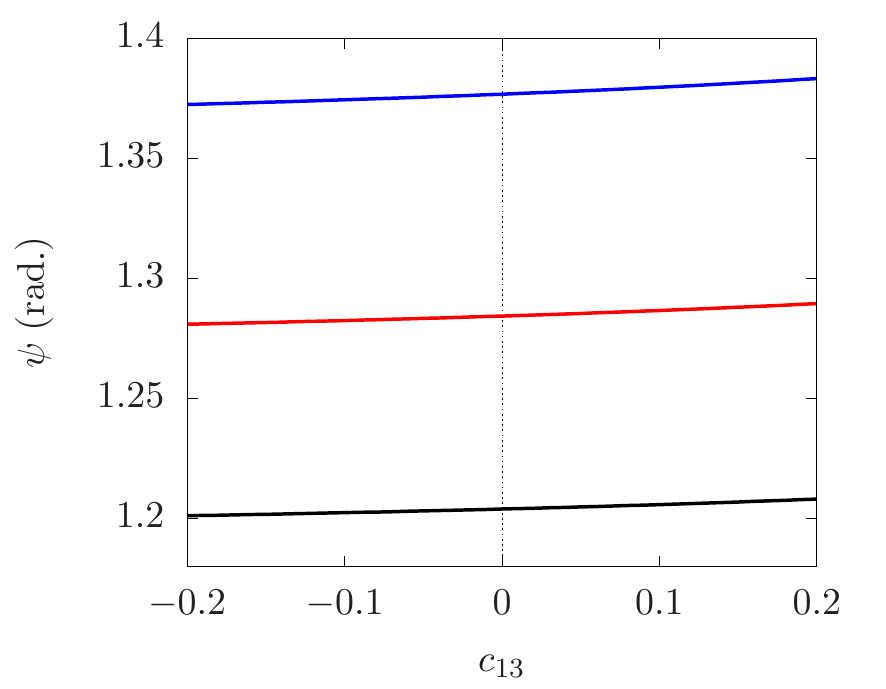}
			\caption{Light deflection of the \AE-black hole of the first uncharged solution. The curves from top to bottom corresponds to bending angle of trajectories with $r_0=5.0M$, $5.2M$, and $5.4M$}.
			\label{fig_lensing1stkind}
		\end{center}
	\end{figure}

	\emph{First solution} --- As mentioned above, in the case of the first-type Einstein-{\AE}ther black hole, the function $G(x)$ is a sixth-degree polynomial. Integrating Eq.~\Eqref{ph_dphidu3} gives the bending angle for a photon which approaches the black hole at the closest (coordinate) distance $r_0=1/x_0$. Fig.~\ref{fig_lensing1stkind} shows the bending angle against $c_{13}$ for some chosen values of $r_0$. We find that the bending angle is enhanced compared to the Schwarzschild case for positive $c_{13}$, and smaller for negative $c_{13}$.
	\begin{figure}
		\begin{center}
			\includegraphics[width=0.48\textwidth]{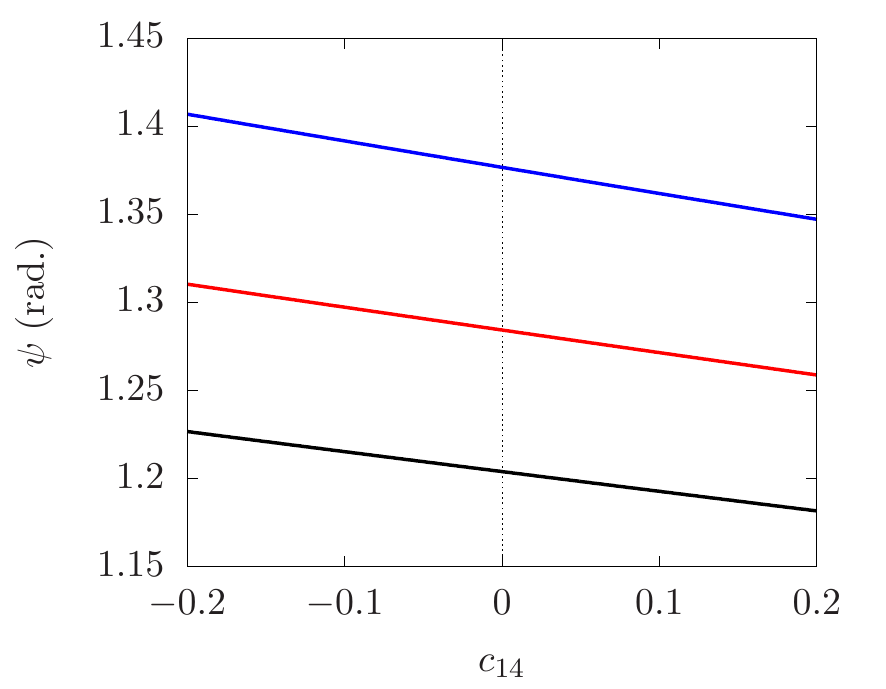}
			\caption{Light deflection of the \AE-black hole of the second uncharged solution $c_{13}=0$. From top to bottom, the coordinate distance of closest approach $r_0$ are $5.0M$, $5.2M$, and $5.4M$.}
			\label{fig_lensing2ndkind}
		\end{center}
	\end{figure}
	
	\emph{Second solution} --- For the second type of Einstein-{\AE}ther black hole, the function $G(x)$ is a fourth-degree polynomial and Eq.~\Eqref{ph_dphidu3} can be solved exactly. More specifically, for the case where $\sigma$ is positive, denote the roots of $G(x)$ as $x_\pm$, $x_0$, and $x_1$. The function $G(x)$ is then factorized in terms of its roots as
	\begin{align}
	G(x)=4M^2\sigma(x-x_1)(x-x_-)(x_0-x)(x_+-x),
	\end{align}
	where, for typical parameter ranges of the black hole, the roots of $G(x)$ have the order 
	\begin{align}
	x_1\leq x_-\leq0\leq x_0\leq x_+.
	\end{align}
	The function $G(x)$ is typically positive for values of $x$ in the range $0\leq x\leq x_0$, and this is the relevant $r=1/x$ for which an incident photon from infinity ($x=0$) approaches the black hole until it reaches a minimum distance $r_0=1/x_0$, before going off to infinity again. 
	The bending angle is then given by (3.147--5 of \cite{gradshteyn2014table})
	\begin{align}
	\psi&=\frac{1}{M\sqrt{\sigma}}\int_0^{x_0}\frac{dx'}{\sqrt{(x-x_1)(x-x_-)(x_0-x)(x_+-x)}}\nonumber\\
	&=\frac{2\mathrm{F}\brac{\zeta,\gamma_1}}{M\sqrt{\sigma (x_+-x_-)(x_0-x_1)}},
	\end{align}
	where $\mathrm{F}(\zeta,\gamma_1)$ is the elliptic function of the first kind with  
	\begin{align}
	\zeta&=\arcsin\sqrt{\frac{(x_+-x_0)x_0}{(x_0-x_-)x_+}}, \label{zeta}\\
	\gamma_1&=\sqrt{\frac{(x_0-x_-)(x_+-x_1)}{(x_+-x_-)(x_0-x_1)}}.
	\end{align}
	
	On the other hand, in the case where $\sigma$ is negative, we have the following order of roots 
	\begin{align}
	x_-\leq0\leq x_0\leq x_+\leq x_1.
	\end{align}
	Again, the function $G(x)$ is typically positive for values of $x$ in the range $0\leq x\leq x_0$. In this case, the bending angle is given by (3.147--3 of \cite{gradshteyn2014table})
	\begin{align}
	\psi&=\frac{1}{M\sqrt{|\sigma|}}\int_0^{x_0}\frac{dx'}{\sqrt{(x-x_-)(x_0-x)(x_+-x)(x_1-x)}}\nonumber\\
	&=\frac{2\mathrm{F}\brac{\zeta,\gamma_2}}{M\sqrt{|\sigma| (x_1-x_0)(x_+-x_-)}},
	\end{align}
	where $\zeta$ is the same as given in Eq.~\Eqref{zeta}, and 
	\begin{align}
	\gamma_2&=\sqrt{\frac{(x_0-x_-)(x_1-x_+)}{(x_+-x_-)(x_1-x_0)}}.
	\end{align}
	Fig.~\ref{fig_lensing2ndkind} shows $\psi$ vs $c_{14}$ (with $c_{13}=0$) for various chosen values of $r_0$. More precisely, the bending angle depends on $\sigma=\frac{2c_{13}-c_{14}}{8(1-c_{13})}$. The bending angle will be larger compared to the Schwarzschild case for positive $\sigma$, and smaller for negative $\sigma$.

	\section{Discussions and Conclusions\label{seccon}}

	In this paper, we study the motion of test particles around two exact charged black-hole solutions in Einstein-{\AE}ther theory. Specifically, we first consider the quasi-periodic oscillations (QPOs) and their resonances generated by the particle moving in the Einstein-{\AE}ther  black hole and then turn to study the periodic orbits of the massive particles. Concerning the study of QPOs we have dropped the usually put-forward assumptions: $\nu_U=\nu_\theta$, $\nu_L=\nu_r$ with $\nu_U/\nu_L=3/2$. We, instead, put-forward a new working ansatz by setting $\nu_U = 2.6(\nu_\theta-\nu_r)$ and $\nu_L =2.6\nu_r$. With this assumption, we have explored in details the effects of the {\ae}ther field on the frequencies of QPOs. We have shown that the value of $r_0$, solution to $\nu_U/\nu_L=3/2$, is much closer to $r_{\text{isco}}$ where the events of accretion and QPOs occur.  This has allowed us to obtain good and complete curve fits for the three microquasars GRO J1655-40, XTE J1550-564 and GRS 1915+105 whether we treat them as static solutions to Einstein-{\AE}ther gravity or as Schwarzschild black holes. 
	
	From the geodesic equations in the two types of Einstein-\AE{}ther black holes, we find that the geodesic equation can be solved analytically for the first type black hole (the black hole with \ae{}ther parameter $c_{14}=0$ but $c_{123}\neq 0$). The innermost stable circular orbits foe two black holes are also analyzed and we find the isco radius increases with increasing $c_{13}$ for the first type black hole while decreases with increasing $c_{14}$ for the second one. We also obtain several periodic orbits and find that they share similar taxonomy schemes as the periodic equatorial orbits in the Schwarzschild/Kerr metrics. These results provide us a possible way to distinguish the two exact charged black holes in Einstein-{\AE}ther theory from the Schwarzschild black hole.
	
	In addition, we have also considered how the radii of circular photon orbits and bending angles for gravitational lensing varies with the parameters of the Einstein-{\AE}ther theory.


	\section*{Acknowledgements}
	
	T.Z. and Q.W. are supported in part by the National Natural Science Foundation of China with the Grants No.11675143, the Zhejiang Provincial Natural Science Foundation of China under Grant No. LY20A050002, and the Fundamental Research Funds for the Provincial Universities of Zhejiang in China under Grants No. RF- A2019015. Y.-K.L is supported by Xiamen University Malaysia Research Fund (Grant no. XMUMRF/2019-C3/IMAT/0007).

\end{document}